# Identification of defects and the origins of surface noise on hydrogen–terminated (100) diamond

Yi-Ying Sung,[1,∇] Lachlan Oberg,[2,3,∇] Rebecca Griffin,[1] Alex K. Schenk,[1] Henry Chandler,[3] Santiago Corujeira Gallo,[3] Alastair Stacey,[4,5] Tetiana Sergeieva,[6] Marcus W. Doherty,[3] Cedric Weber,[3] and Christopher I. Pakes,[1]

[1] Department of Mathematical and Physical Sciences, La Trobe University, Bundoora, Victoria 3086, Australia.

[2] The Australian National University, Canberra, ACT 2600, Australia.

[3] Quantum Brilliance Pty. Ltd., 60 Mills Road, Acton, ACT 2601, Australia.

[4] School of Science, RMIT University, Melbourne, VIC 3000, Australia.

[5] Princeton Plasma Physics Laboratory, Princeton University, Princeton, New Jersey 08540, USA.

[6] Quantum Brilliance GmbH, Colorado Tower, Industriestraβe 4, 5.OG 70565 Stuttgart, Germany.

[∇] These authors contributed equally to this work.

## ABSTRACT

Near-surface nitrogen-vacancy centres are critical to many diamond-based quantum technologies such as information processors and nanosensors. Surface defects play an important role in the design and performance of these devices. The targeted creation of defects is central to proposed bottom-up approaches to nanofabrication of quantum diamond processors, and uncontrolled surface defects may generate noise and charge trapping which degrade shallow NV device performance. Surface preparation protocols may be able to control the production of desired defects and eliminate unwanted defects, but only if their atomic structure can first be conclusively identified. This work uses a combination of scanning tunnelling microscopy (STM) imaging and first-principles simulations to identify several surface defects on H:C(100)-2×1 surfaces prepared using chemical vapour deposition (CVD). The atomic structure of these defects is elucidated, from which the microscopic origins of magnetic noise and charge trapping is determined based on modelling of their paramagnetic properties and acceptor states. Rudimentary control of these deleterious properties is demonstrated through STM tip-induced manipulation of the defect structure. Furthermore, the results validate accepted models for CVD diamond growth by identifying key adsorbates responsible for nucleation of new layers.

## KEYWORDS

Diamond, Scanning Tunnelling Microscopy, Surface Defects, Chemical vapour Deposition, Quantum

# INTRODUCTION

Many diamond-based quantum technologies exploit the negative charge state of the nitrogen-vacancy (NV$^-$) centre [1]. This colour centre in diamond has a unique combination of optical and spin properties which enable room-temperature quantum information processing (QIP) [2, 3] and quantum sensing [4]. Near-surface NV centres are often a critical component of device architectures pursuing these applications. For example, QIP schemes may require near-surface NV centres for compatibility with surface control structures such as electrodes for implementing qubit initialisation and readout. Similarly, nanosensors often require near-surface NV centres for maximising sensitivity to external field sources [5].

Realising the full capabilities of diamond quantum technologies which utilise near-surface defect centres requires the preparation of high-quality surfaces for two key reasons. Firstly, defects on the diamond surface are known to degrade the performance of engineered QIP and sensing devices which rely on near-surface NV centres [6–8]. Unpaired electron spins, sometimes associated with surface dangling bonds (DBs), can introduce dephasing and reduce NV$^-$ coherence times [6, 7, 9, 10]. Furthermore, surface defects with mid-gap acceptor states act as charge traps which cause electrical and magnetic noise and screen electric fields in *n*-type samples [8, 11, 12].

Secondly, ideal QIP device schemes require the deterministic placement of an array of NV centres with nanometre precision [13]. This has motivated the development of bottom-up, atom-scale approaches to diamond device nanofabrication using precision hydrogen depassivation lithography (HDL) [14] to create reactive surface sites and followed by targeted nitrogen chemistry [15, 16]. Any such fabrication protocol requires, as a starting point, an atomically smooth and ordered surface platform with a low concentration of native surface defects. Furthermore, these bottom-up fabrication protocols require engineering of targeted surface defects (specifically chemisorption sites) via HDL [17]. It is therefore imperative, both for device performance and precision fabrication, to develop surface preparation protocols which target and eliminate unwanted defects both in as-grown diamond substrates and in engineered devices. Importantly, the identification of diamond surface defects is a vital pre-requisite for informing these protocols.

The characterisation of diamond surface defects has therefore been the focus of several recent experimental and theoretical works. Experimental studies have largely employed widefield spectroscopic techniques to classify their aggregate features including defect density, bonding hybridisation (e.g. sp$^2$ versus sp$^3$), and energy distributions [8, 9, 18, 19]. First-principles simulations have subsequently identified defects that align with these experimental observations. For example, Stacey *et al*. [8] used a combination of near edge X-ray absorption fine structure spectroscopy and density functional theory (DFT) to detect dozens of defects which could be responsible for magnetic noise and charge trapping at the diamond surface. Unfortunately, this approach cannot conclusively identify which defects are present on the surface. Widefield spectroscopy simply lacks the spatial resolution to probe the atomic structure of the defects. This motivates the use of scanning tunnelling microscopy (STM) which offers atomic-scale imaging and the capability for single-defect spectroscopy.

Previous ultra-high vacuum (UHV) STM studies on diamond have underscored the challenge of imaging defects on common diamond surfaces [20]. These challenges include low surface conductivities, the material hardness of the surface, a lack of in-situ surface processing and tip-conditioning protocols, and high surface roughness contributing to tip destabilization. In contrast to progress made on silicon and metal surfaces, the combination of these effects has historically hindered STM imaging of diamond with atom-scale resolution. Several defects including DBs, depassivated dimers [14], carbon vacancies, and anti-phase boundaries [21], have been tentatively identified on (100) surfaces. However, these assignments are not conclusive as STM topographies may not trivially reflect the real-space atomic surface structure. Moreover, many other defects observed on these surfaces remain unidentified to date. First-principles simulations of STM topographies can assist in defect identification through comparison with experimental results [22, 23], but have been underutilised on diamond. Moreover, many existing simulation frameworks fail to account for important electrostatic effects which are inherent to wide-band gap semiconductors during imaging (e.g., band bending). These effects can have a significant influence on the observed STM topography and thereby further complicate their interpretation [24, 25].

This work overcomes these barriers to attain unprecedented insight into the surface defects relevant to near-surface quantum devices. This is accomplished using a combination of high-resolution STM imaging and state-of-the-art first-principles simulations which account for tip-induced surface effects. Ultra-high vacuum STM is used to visualise several defects on H:C(100)-2×1 surfaces grown by chemical vapour deposition (CVD). The atomic structure of these defects is then determined by comparing the measured STM topographies with those obtained using multi-scale STM simulations. These simulations are termed multi-scale because they combine mesoscale electrostatic modelling with nanoscale DFT. The electronic and magnetic structure of the defects are then classified based on symmetry observed in the STM topographies. This reveals the origins of paramagnetic noise and charge trapping on the diamond surface and validates accepted models for the chemistry of diamond CVD overgrowth. Finally, we demonstrate rudimentary control over these magnetic and trapping properties through tip-induced modifications.

While this work focuses solely on the H-terminated diamond surface, which is relevant to proposed strategies for atom-scale lithography of QIP devices, we note that H-termination is not necessarily desirable in engineered devices for all quantum applications[26,27]. The negative electron affinity causes band bending which can destabilise the negative charge state of near-surface NV centres. Other terminations, such as fluorine (F) or oxygen (O), are considered more suitable for quantum sensing because they possess a positive electron affinity [28]. However, STM capabilities have yet to be substantively explored on these alternative surfaces. Consequently, while this study considers H:C(100)-2×1, the simulation methodology and analysis framework can easily be adapted to other surfaces once the technological and engineering capabilities are further developed. The results may even be directly transferable. For example, F-terminated diamond possesses a similar surface reconstruction and bonding chemistry to H-terminated diamond [29], and therefore similar defects may be expected.

## RESULTS AND DISCUSSION

**Symmetry classification of observed defects on H:C(100)-2×1.** Figure 1 shows empty-state topographic STM images of the clean H:C(100)-2 × 1 surface following annealing of the sample after insertion into UHV (see Experimental Section). This provides a broad overview of several different point defects visible on the surface. The defects are scattered sporadically amongst the characteristic dimer rows, step edges, and terraces aligned at right angles as typically observed for a 2×1 surface reconstruction. For reference, these dimer rows consist of σ-bonded CH–CH pairs which form a linear bright feature visible in the STM topography. Comparison of images obtained for the same sample after annealing in-situ at two different temperatures (723K and 823K) suggests that both the number and nature of the defects observed changes depending on the annealing temperature.

We have classified the observed defects into four different symmetry classes based on the qualitative features in their associated STM topographies, labelled as $1s^+$, $1s^-$, $2s$, and $2p$. These symmetry classes have been established through a systematic analysis of simulated topographies for 25 different defects. The topographies were produced using multi-scale modelling and have been compiled into an STM defect library as described in the Supplementary Information (SI). Note that the nomenclature used refers explicitly to the symmetry of the STM topography and is not necessarily representative of the defect's orbital structure.

However, as discussed later, the symmetry class does reflect the underlying electronic and magnetic structure of the defects.

The $1s^+$ defects are characterised by a singular bright feature as indicated in Figure 1(b). This feature originates from occupied orbitals which are both localised to the defect and possess one-electron energies near the surface valence band maximum (VBM). Analysis of the electronic and magnetic structure of the observed defects (as predicted by DFT) is presented within the SI. The $1s^-$ defects are characterised by a prominent singular dark feature, as indicated in Figure 1(a), indicating an absence of surface-related orbitals local to a vacancy. The 2s defects are characterised by a 'doughnut' shaped feature, as shown in Figure 1(a), consisting of a central bright spot surrounded by a dark ring reflecting the absence of any localised orbitals near the VBM. Finally, the 2p defects, observed in Figure 1(b), are characterised by two bright features separated by a nodal plane again produced by occupied orbitals near the VBM.

Referring to Figure 1(a), after annealing at a temperature of 723 K, the majority of defects observed (approximately 75%) have clear 2s symmetry. Most of these are positioned within the centre of the dimer rows, with only a few exceptions that appear off-centre (shown in SI5). Furthermore, about 90% of the observed 2s defects are identical in size. This indicates that they are likely the same defect, with minor variations due to variation in tip properties during scanning. Of the remaining defects, most display a dark singular feature characteristic of $1s^-$. Physisorbed or weakly bonded impurities (such as adsorbates from the atmosphere) should be eliminated by the annealing process. Therefore, as elucidated below, the 2s and $1s^-$ defects are expected to be stable by-products of the CVD growth process.

Upon increasing the anneal temperature to 823 K, Figure 1(b) reveals a significant increase in the number of bright features characteristic of $1s^+$ and 2p defects, along with a reduction in the number of 2s defects. The $1s^+$ and 2p defects are presumed to form via thermal desorption of hydrogen (H) atoms from the dimer rows and from 2s defects, respectively, thereby creating DBs which produce bright features in the STM topography on account of the occupied orbitals. Temperatures above 1100 K are typically required to induce widespread H desorption on the diamond surface [30]. However, limited amounts of desorption may occur below this temperature. Furthermore, based on the prevalence of bright features, it appears that H atoms on defect sites are more susceptible to thermal desorption than on dimer rows. This suggests that a lower energy barrier is required for desorption.

**Identification of point defects on H:C(100)-2×1.** High-resolution empty-state STM images were used to determine the atomic structure of several commonly occurring surface defects. The imaged defects have been identified through comparison of their experimental topographies with simulations obtained using the multi-scale method (refer to Figure SI3 for the full defect library). This comparison has been made based on qualitative and quantitative features of the topographies, though no explicit fitting procedure has been performed. A side-by-side comparison of the experimental and simulated topographies are presented in Figures 2 and 3; these include one-dimensional line profiles sampled through the defect centre points. While multiple instances of each defect were found on the surface, only one representative experimental image has been illustrated in each case. Moreover, only the most prevalent defects on the surface have been identified. We note that there are other defects which could not be clearly identified, but these were far less prevalent.

Defects from each symmetry class are now considered in turn. Beginning with the $1s^+$ defects, two examples have been identified as shown in Figure 2(a): a bare C–C dimer and an $sp^3$ DB (i.e. a single de-passivated C atom). These can be recognised by a distinctive bright feature originating from occupied orbitals near the surface VBM. While qualitatively the appearance of these defects is the same, the C–C and $sp^3$ DB can be differentiated by their relative widths and position with respect to the centre of the underlying dimer rows.

This is demonstrated within Section 3 of the SI. The simulated images provide an excellent quantitative description of both the feature height and width. Moreover, note that this agreement requires no additional fitting of the simulations with the same parameters of the multi-scale model being used for both defects. As discussed above, the significantly higher concentration of C–C dimers and $sp^3$ DBs on the surface following the higher temperature anneal is consistent with these defects being produced via desorption of H atoms from the dimer rows.

Two different $1s^-$ defects were identified by their characteristic dark features: H-passivated C and $C_2$ vacancies. As illustrated in Figure 2(b), these two defects can be differentiated by their width and position relative to the centre of the dimer row. The dark features indicate an absence of surface-related orbitals local to the vacancy; moreover, due to the absence of any bright features, they are not suspected to be the unsaturated vacancies ((xii) and (xiii) of Figure SI3). Saturated vacancies are an expected by-product of diamond CVD growth due to abundance of hydrogen radicals in the CVD plasma [31]. This is consistent with their presence on the surface of as-grown samples following a low temperature anneal.

The simulated topographies of the $1s^-$ bare strong qualitative similarity to the experimental images. Analysis of the line profiles reveal that the widths of the dark features are well-captured by the multi-scale model. However, the simulations overestimate the depth of the dark regions; this is particularly severe for the $C_2$ vacancy and exceeds the experimental depth by over 1 Å. The overestimation of dark features occurs consistently throughout the simulations and will similarly be encountered for both 2s and 2p defects. As discussed in greater detail within the SI, the source of this discrepancy is expected to be the bluntness of experimental STM tips. The simulations assume that all tunnelling current flows through a single atom at the tip apex This situation is likely unphysical, as a realistic tip apex would be composed of a blunt cluster of atoms. The lateral footprint of the real tip means that the tunnelling current drawn will be higher than simulated when moving into a depression. Due to the exponential relationship between tunnelling current and distance, dark features are therefore overestimated in the simulations.

Next consider the 2p and 2s defects illustrated in Figure 3. Two different 2p defects were identified as illustrated in Figure 3(a). These are the CH bridge and the $CH_2$ adsorbate. For both defects, the central C atom is $sp^2$ hybridised and forms covalent bonds with H atoms and surface C atoms. The final unpaired electron forms an unhybridised 2p orbital with a nodal plane aligned either perpendicular or parallel to the dimer rows. Although the simulated topographies are in qualitative agreement with experiment, the line profiles reveal quantitative discrepancies. For the CH bridge, the simulated depth of the dark feature associated with the nodal plane is overestimated. This is again consistent with a blunt apex on the experimental tip, especially given the small spatial width of the nodal plane. For the $CH_2$ adsorbate both the width and height of the bright features are poorly reproduced by simulation. It seems doubtful that this could be attributed to the bluntness of the tip. Hence, it is plausible that this defect may be incorrectly identified as a $CH_2$ adsorbate, despite the qualitative similarities of the STM topographies. However, the fact that these 2p defects appear after a high temperature sample anneal suggests that both the CH bridge and $CH_2$ adsorbate may be present on the surface due to thermal desorption of H atoms from a $CH_2$ bridge and $CH_3$ adsorbate respectively.

Finally, three different 2s defects were identified by their distinctive 'doughnut' features. As depicted in Figure 3(b), these are the $CH_3$ adsorbate, the $CH_2$ bridge (dimer), and the $C_2H_2$ bridge (dimer). Each defect is differentiated by the relative size and positioning of its dark ring. The $CH_3$ adsorbate and $CH_2$ bridge produce circular rings positioned at either the edge or centre of a dimer row respectively. The $C_2H_2$ dimer has an elliptic shape located in the centre of the dimer.

**Atomic-scale validation of CVD models** The observed 2s defects provide a remarkable validation of accepted models for (100) diamond CVD growth [31]. These models have previously been established through *ab initio* simulations [32], mesoscale modelling of growth rates and surface morphology [33,34], and plasma experiments and simulation [35]. The STM analysis presented here provides further experimental validation of this model at the atomic scale. In particular, all three of the 2s defects observed correspond to a specific stage in the CVD diamond growth process. Firstly, a $CH_3$ defect is produced through adsorption of an impinging methyl radical from the CVD plasma. This $CH_3$ defect incorporates into the dimer to form a $CH_2$ bridge which subsequently acts as a nucleation point for new layer growth. The formation of this new layer begins through adsorption of another methyl radical which incorporates to produce a $C_2H_2$ adsorbate. This process is depicted graphically through chemical structure diagrams in Figures SI9 and SI10 of the SI. Recall from the preceding section that most defects observed in the wide-area scan (Figure 1) possess 2s symmetry. The most prevalent of these can be identified as the $CH_2$ bridge, thereby validating its essential role of nucleation during CVD growth on the (100) surface.

**Identified defects as potential noise sources.** Having established the capability to identify atom-scale surface defects, we now consider the role that each class of identified defect may have in diminishing the performance of engineered QIP and sensing devices. Defects with either paramagnetic DBs or mid-gap acceptor states are generally undesirable for quantum devices utilising near-surface NV centres. In the former case, DBs can introduce magnetic noise which dephases the $NV^-$ electronic spin. In the latter case, mid-gap states may act as charge traps in *n*-type samples which also produce magnetic noise along with electrical noise and screening of external electric fields [12,36].

First-principles modelling reveals that none of 2s defects identified on the as-grown H:C(100)-2×1 surface are paramagnetic or possess mid-gap acceptor states. This modelling is presented within Section 2.3 of the SI. From analysis of the wide-area STM images following the low-temperature anneal, over 75% of the observed defects exhibit 2s symmetry. Due to their positioning within the centre of the dimer rows, the majority of these defects are suspected to be the $CH_2$ bridge. For the samples we have prepared, we therefore conclude that most of the observed defects would not adversely affect the properties of near-surface NVs relevant to QIP and quantum sensing.

Similarly, none of the $1s^-$ defects identified on the surface are expected to negatively impact quantum applications involving NV centres. As depicted in the STM defect library (Figure SI3 of the SI), $1s^-$ vacancies may produce mid-gap states if the surrounding carbon atoms form double bonds (C=C). However, the absence of any bright features associated with the $1s^-$ vacancies identified in Figure 2 (b) indicate that they are saturated. Hence, no mid-gap states are expected.

In contrast, both the $1s^+$ and 2p defects identified on the surface are expected to negatively impact the performance of NV-based quantum devices. The $sp^3$ DB is paramagnetic and both 2p defects similarly possess a paramagnetic DB. While the C=C dimer is not paramagnetic, it possesses an acceptor state (the anti-bonding orbital) approximately 4 eV above the valence band maximum. This is expected to act as a charge trap in *n*-type samples, which are typically used for both QIP and sensing. Analysis of the wide-area STM images in Figure 1 indicates that 0.5%–2.0% of surface C sites on the as-grown surface possess a $1s^+$ or 2p defect. The majority of these defects are $sp^3$ DBs and C=C dimers.

**Tip-based manipulation of defect properties.** Finally, we consider approaches to alter surface defects in a targeted fashion. Rudimentary control over defect structure was achieved through sample voltage sweeping with the STM tip held in the vicinity of a defect. Figure 4 presents three different examples of defect modification induced by ramping the sample bias from -2.0 V to +2.5 V over a duration of 3 seconds, leading to either tip-induced H-atom desorption from a defect or passivation of defect bonds with an additional H-atom (termed capping). In each case, Figure 4 illustrates STM images of the same defect presented before and after the voltage sweep. The resulting change in topographical features provides clear evidence of structural modifications.

Figure 4(a,b) illustrates the transformation of a 2s defect into a 2p defect following a voltage sweep. The 2s defect can be identified as the $CH_2$ bridge (dimer), whereas the identification of the 2p defect is more speculative. The two lobes are oriented perpendicular to the dimer row which is consistent with the $CH_2$ adsorbate. However, there is no asymmetry to the corrugation profile like that observed in Figure 3 (b). The exact identity of the defect is therefore uncertain, although it has clear 2p symmetry. Regardless, the transformation of a 2s to a 2p defect indicates that an unhybridized 2p orbital has been formed through tip-induced H-desorption. Figure 4(a,b) therefore presents a demonstration of controlled single-atom HDL constrained to $CH_2$ defects. The desorption achieved was of low yield and challenging to replicate, occurring on defects situated up to 1 nm away from the tip apex. Hence, this process is not yet suited to atom-scale fabrication, which requires controlled depassivation lithography on passivated terraces with atomic-scale precision. Further work is therefore required to extend the methods developed here to achieve this goal. Specifically, we note that that desorption on both $CH_2$ defects and terraces are expected to occur through a similar mechanism; localised bond heating induced by inelastic tunnelling [37].

Figures 4(c,d) and 4(e,f) illustrate examples of H atom capping. Hydrogen capping is commonly used for error correction during HDL on silicon. It entails passivating erroneous dangling bonds using a H-functionalised tip [38]. Note that functionalised tips were not deliberately prepared in this study. Instead, we attribute the presence of H as a by-product of earlier tip-sample bias sweeps. Regardless, H-capping of multiple $sp^3$ dangling bonds is evident in Figure 4(c,d). The bright features ($1s^+$) are effectively erased, and the resulting passivated dimers become indistinguishable from the rest of the terrace. Another example of capping depicted in Figure 4(e,f) is the transformation of a CH bridge into a $CH_2$ bridge.

The tip-induced capping of diamond surface defects with a H atom may be used to modify the class of defect and therefore engineer its electronic and magnetic properties. In particular, H passivation of $1s^+$ and 2p defects can eliminate the paramagnetic and charge trapping behaviour which is detrimental to shallow NV centres. While we have not yet achieved deterministic control of defect structure, these demonstrations offer a pathway to atomic-scale defect engineering at the diamond surface. In combination with the defect library -- which enables clear evaluation of the results of defect modification -- more comprehensive studies exploring the dynamics of these processes and the influencing variables are now possible. This may ultimately lead to deterministic control over defect properties using STM manipulation. Possible applications include the bespoke design of atomic-scale reaction schemes for complex surface engineering, STM-based patterning of specific sites, and enhancing the sensitivity and performance of diamond quantum devices.

CONCLUSION

Through a combination of STM imaging and first-principles simulations, the atomic structure of several point defects native to the H-terminated surface of (100) diamond have been identified. The electronic and magnetic structure of the defects have been classified based on the symmetry observed in their STM topographies. Over 75% of the observed defects exhibit 2s symmetry, the majority of which are suspected to be $CH_2$ bridge defects, a by-product of CVD growth. Less than 2% of the surface C sites possess $1s^+$ or 2p symmetry which are expected to introduce magnetic noise, electrical noise, or electric field screening that would be detrimental to QIP or quantum sensing schemes that rely on shallow NV centres. Tip-induced modification of defects via single H-atom desorption and adsorption (capping) has been demonstrated as a means of engineering the properties of isolated defects, offering a route to improving the performance of near-surface diamond quantum devices.

The identification and characterisation of defects observed by STM achieved in this work will inform the development of future bottom-up, atom-scale lithography protocols for engineering QIP devices in diamond. Similar STM and multi-scale modelling analysis may also be applied to other diamond surface terminations such as fluorine or oxygen, where similar associations between the observed symmetry of STM features and defect electronic structure are to be expected.

## EXPERIMENTAL SECTION

**Diamond sample preparation.** The sample preparation procedures and data acquisition were performed in two separate vacuum chambers (CVD and STM). A type IIa single-crystal diamond with a (100) orientation, purchased from Element Six Ltd, was used as a substrate for the CVD growth of a single-crystal surface layer. The sample was pre-treated by exposure to a $H_2$-2%$O_2$ plasma (1.2 kW) at gas pressure of 133.3 mbar for 2 hours to clean and etch the surface. The new diamond layer was then grown from a $H_2$-1.3%$CH_4$ gas mixture at 120.0 mbar with a plasma power of 1.1 kW for 1 hour. Finally, a boron-doped epilayer of 2.5 μm thickness was produced by flowing trimethylboron (2% TMB in H2) into the chamber at a flow rate of 0.15 sccm in the $H_2$-1%$CH_4$ gas mixture (B/C ~ 1000 ppm).

The H-termination was performed in a Seki 6300 CVD reactor, using a lower $CH_4$% concentration (0.2% $CH_4$/$H_2$) but the same B/C ratio. The power, pressure and temperature were similar to the stated growth conditions. After 5 minutes, the B and C sources were turned off and the plasma was gradually turned down in power and pressure over 2 minutes, during which time the expected CH4 and TMB concentration is expected to drop by more than a factor of 10, and it was turned off at 26 mbar and 600W. Both $H_2$ and $CH_4$ go through gas purifiers to remove impurities. The CVD reactor is a Seki 6300.

The sample was then transferred to ambient conditions, mounted on a Mo sample holder, and inserted into the UHV STM chamber. Since the H-terminated diamond surface is very stable in air, the sample only requires annealing in vacuum to typically 723 K to remove water adsorbates and airborne contaminants [39]; the sample anneal temperature was estimated by using an IR pyrometer to monitor the temperature of the Mo sample holder, as diamond does not emit in the infrared range.

**Scanning tunnelling microscopy.** STM measurements were performed using a SPECS Aarhus STM with tungsten tips purchased from SPECS, operating at room temperature and under UHV conditions with a base pressure of $5 \times 10^{-10}$ mbar. Image and data acquisition were undertaken using a Nanonis STM controller. Constant-current STM imaging was performed with a sample bias between 2.0 V and 3.0 V, and with different set currents depending on the conditions of the tip. Tunnelling therefore occurs between occupied states of the tip and unoccupied states of the sample.

**Multi-scale STM simulations.** The multi-scale STM simulations integrated electrostatic effects from mesoscale modelling into DFT. Substantial details regarding these simulations is presented in Section 1 of the SI and references therein.

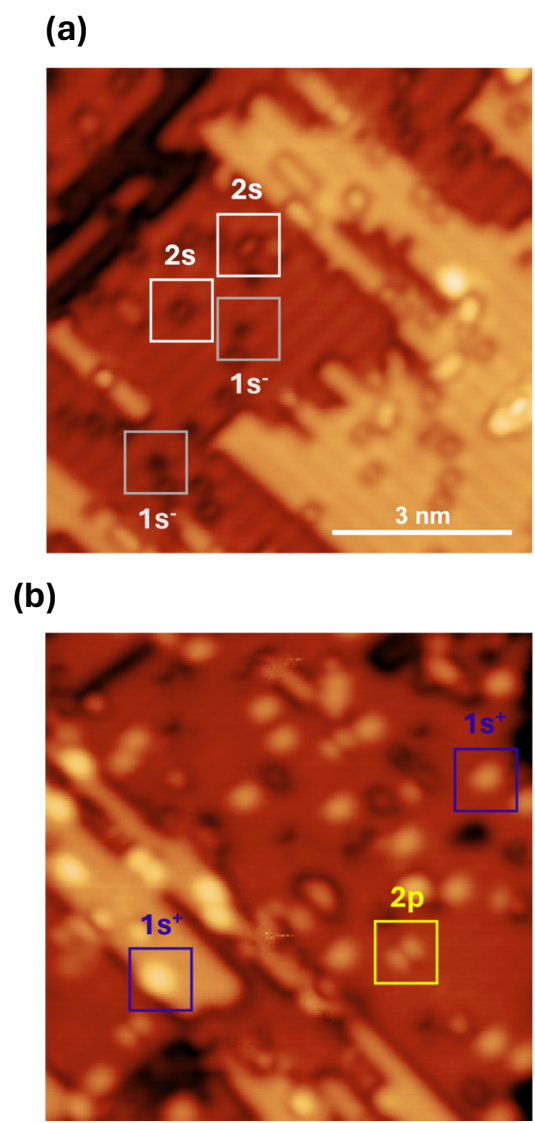

**Figure 1**. STM topographic images illustrating several types of defect observed on H:C(100)-2 × 1 after annealing the sample in UHV at (a) 723 K and (b) 823K. Refer to text for classification of labelled defects. STM tunnelling parameters: (a) V = 2.5 V; I = 200 pA; (b) V = 2.0 V; I = 200 pA.

**(a)**      **1s⁺ symmetry defects**

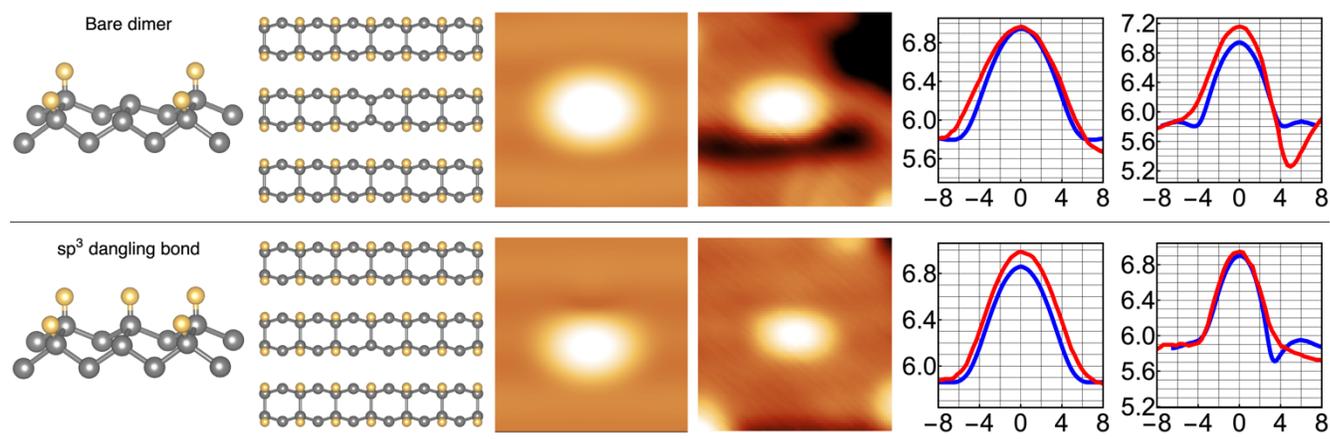

**(b)**      **1s⁻ symmetry defects**

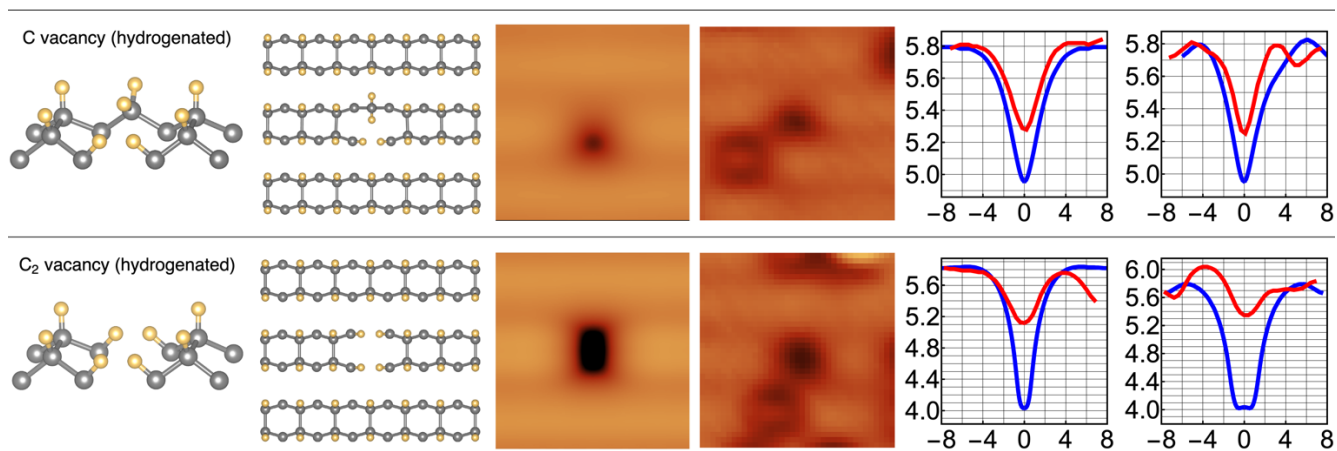

**Figure 2**. STM-based identification of defects with 1s⁺ and 1s⁻ symmetry observed on the H:C(100) – 2 × 1 surface. From left to right, in each case, are the defect geometries depicted from side and top profiles, simulated STM image, experimental STM image, and simulated (blue) /experimental (red) line profiles sampled through the centre of each defect. The horizontal axis for the line profile indicates the distance either parallel ($x$) or perpendicular ($y$) to the dimer row. The vertical axis is the tip height during constant current imaging. All units are in Angstroms.

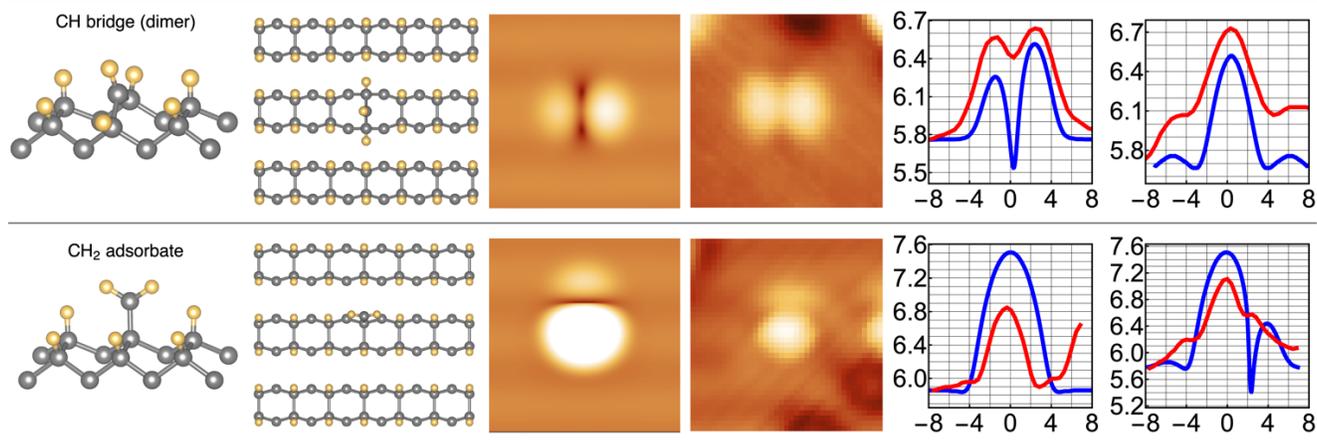

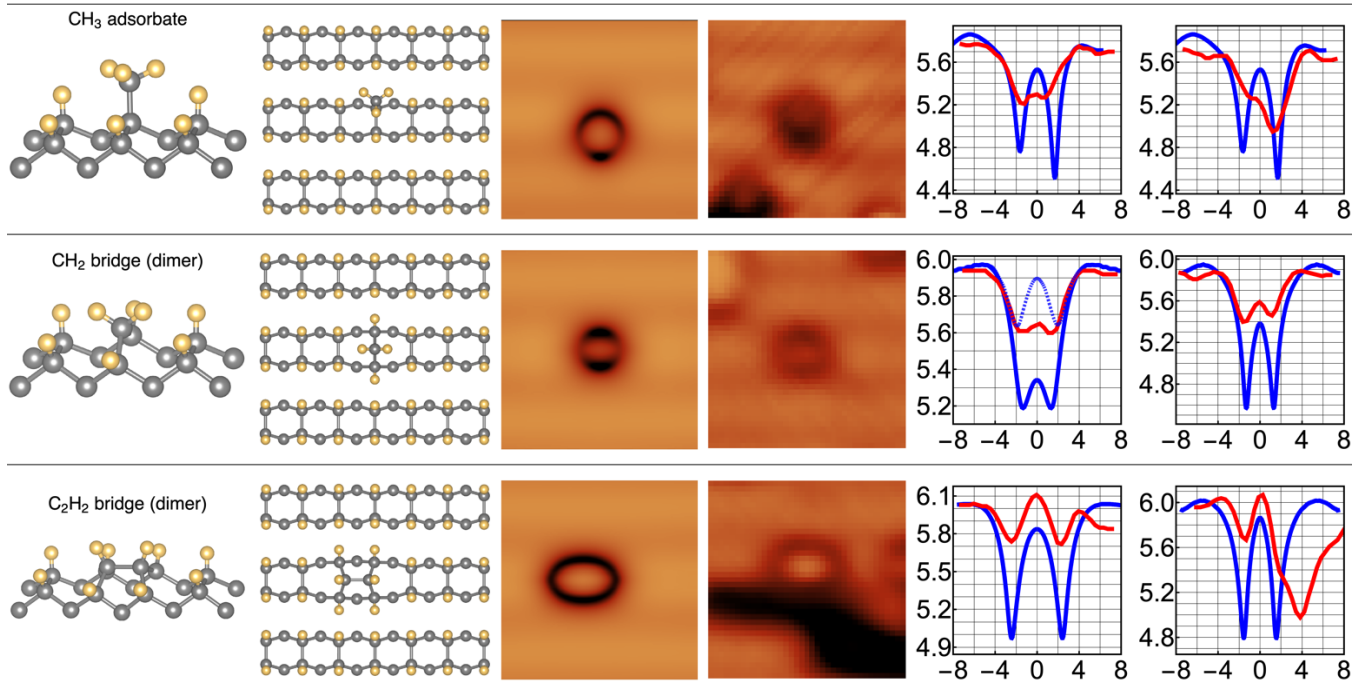

**Figure 3**. STM-based identification of defects with 2p and 2s symmetry observed on the H:C(100) – 2 × 1 surface. From left to right, in each case, are the defect geometries depicted from side and top profiles, simulated STM image, experimental STM image, and simulated (blue) /experimental (red) line profiles sampled through the centre of each defect. The horizontal axis for the line profile indicates the distance either parallel (*x*) or perpendicular (*y*) to the dimer row. The vertical axis is the tip height during constant current imaging. All units are in Angstroms. The blue dashed line in the x line profile for the CH$_2$ bridge dimer is obtained by averaging the tunnelling current within a sphere of radius 1 Å about the tip apex. This averaging process replicates the bluntness of the tip apex and reduces the discrepancy between experimental and simulated line profiles. Further details are provided in the SI.

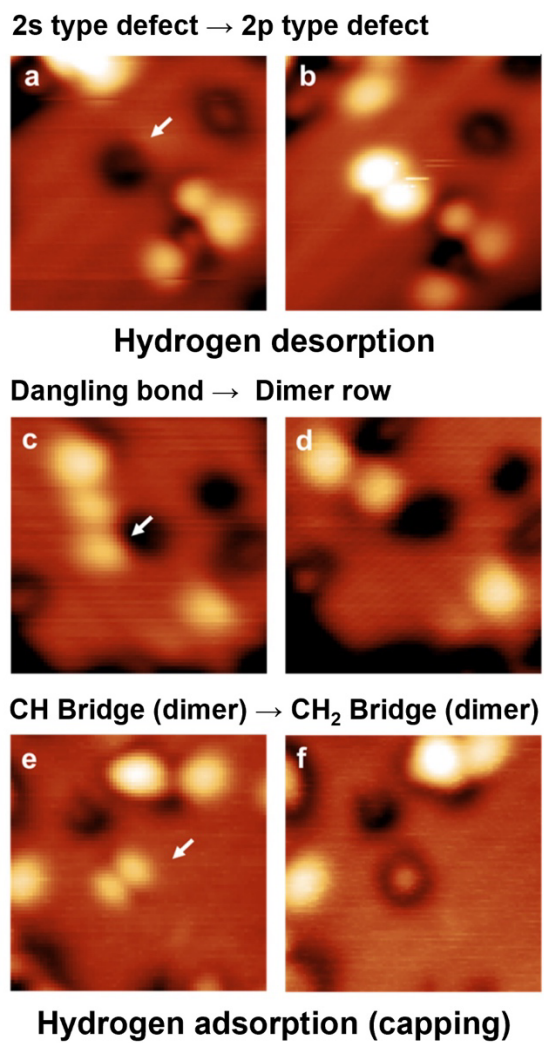

**Figure 4**. STM images illustrating examples of tip-induced surface defect modification induced by voltage sweeping. Each pair of images illustrates changes in defect structure before (left) and after (right) the sample bias is ramped to 2.5 V. Conversion of a CH2 bridge to a CH bridge via H desorption (a, b); conversion of a dangling bond to a dimer via H-adsorption (c, d); conversion of a CH bridge to a CH2 bridge via H-adsorption (e, f). Arrows pinpoint the location of a surface tip-induced reaction.

## Acknowledgements

This research was supported by the Australian Government through the Australian Research Council's Linkage Project funding scheme (project LP200301428). This work was performed in part at the Melbourne Centre for Nanofabrication (MCN) in the Victorian Node of the Australian National Fabrication Facility (ANFF). This research was undertaken with the assistance of resources and services from the National Computational Infrastructure (NCI), which is supported by the Australian Government.

# Identification of defects and the origins of surface noise on hydrogen–terminated (100) diamond

Yi-Ying Sung,[1,∇] Lachlan Oberg,[2,3,∇] Rebecca Griffin,[1] Alex K. Schenk,[1] Henry Chandler,[3] Santiago Corujeira Gallo,[3] Alastair Stacey,[4,5] Tetiana Sergeieva,[6] Marcus W. Doherty,[3] Cedric Weber,[3] and Christopher I. Pakes,[1]

[1] Department of Mathematical and Physical Sciences, La Trobe University, Bundoora, Victoria 3086, Australia.

[2] The Australian National University, Canberra, ACT 2600, Australia.

[3] Quantum Brilliance Pty. Ltd., 60 Mills Road, Acton, ACT 2601, Australia.

[4] School of Science, RMIT University, Melbourne, VIC 3000, Australia.

[5] Princeton Plasma Physics Laboratory, Princeton University, Princeton, New Jersey 08540, USA.

[6] Quantum Brilliance GmbH, Colorado Tower, Industriestraβe 4, 5.OG 70565 Stuttgart, Germany.

[∇] These authors contributed equally to this work.

(Dated: 30 May, 2024.)

Section 1 discusses the context, theory, and implementation of the multi-scale STM model. Section 2 presents the STM defect library on H–C(100): $2 \times 1$ produced using the multi-scale model. The overestimation of dark features in simulated images is also discussed and the electronic band structures of the observed defects are presented. Section 3 demonstrates additional examples of defects identified in the STM experiments. These highlight the methods used to differentiate between similar looking defects. Section 4 provides reaction diagrams of the accepted CVD growth process.

## 1. Multi-scale STM simulations

### 1.1 Context

Diamond is an archetypal insulator with an indirect bandgap of 5.47 eV. As a result, there is effectively no intrinsic carrier conduction at room temperature. P-type conductivity may be realised via substitutional boron doping which introduces an acceptor energy at 0.37 eV above the valence band maximum (VBM) [1]. Boron is therefore classified as a deep acceptor with incomplete ionization at room temperature. This introduces a host of electrostatic screening effects which must be considered in STM simulations of boron-doped diamond [2].

One of the most important effects is band bending, which results in a voltage drop over the scale of tens of nanometres to microns beneath the surface. This is due to screening of the tip-induced electric field by ionised boron dopants or mobile holes. Consequently, the voltage drop ($V_d$) between the tip and the diamond surface is reduced relative to the applied bias ($V_b$). The energies of surface-related states are therefore translated by a reduced amount $V_d \leq V_b$ relative to the tip Fermi level under applied bias.

Note that band bending is typically considered negligible during STM on metals and some semiconductors with shallow donors/acceptors. In these cases, the surfaces of the tip and sample can be conceptualised as an ideal capacitor. The voltage drop is then roughly equivalent to the applied bias ($V_d = V_b$). This is not necessarily the case for boron-doped diamond, which displays a reduced capacitative response. Instead, careful consideration of band bending is required to correctly align the tip Fermi level with the sample's surface states under an applied bias. This alignment is extremely important during STM because it determines which surface states contribute to tunnelling.

In addition to band bending, there are several more electrostatic effects relevant to STM on boron-doped diamond. These include modifications to the surface wavefunctions due to the large electric field within the vacuum region between tip and sample, as well as mixing of bulk-like and surface states due to the internal potential produced by band bending.

This work employs the multi-scale model developed in Chapter 4 of [3] to incorporate these mesoscale electrostatic effects into nanoscale STM simulations. Extensive details regarding the multi-scale model have previously been presented in [3], including derivations of the theoretical approach and validation of its applicability to diamond STM. While the full details are not reproduced in this supplementary, the key components of the model will now be discussed. These are electrostatic modelling, first-principles calculations, and simulation of the STM topographies.

The electrostatic modelling entails full characterisation of the screening effects which occur in the combined tip and sample system. Finite element analysis is used to determine the electrostatic potential throughout the sample and vacuum region for any given experimental conditions. These conditions include the applied bias ($V_b$), dopant density, temperature, and tip dimensions.

These electrostatic effects are then integrated into first-principles modelling of the H–C(100):2 × 1 surface. Density functional theory (DFT) is used to determine the wavefunctions and energies of the defect and surrounding surface. This electronic solution is then modified by the electrostatic potential obtained from the finite element analysis. Namely, the screening effects are systematically integrated through both self-consistent and perturbative techniques.

Finally, the modified wavefunctions and energies are used to simulate STM topographies. This is performed in the conventional manner using Bardeen's perturbation theory to calculate the tunnelling matrix elements [4]. Each component of this model is now expanded in further detail.

## 1.2 Key elements of the model

### 1.2.1 Electrostatic modelling

Electrostatic modelling of the combined tip and sample system was performed using COMSOL Multiphysics (version 5.3). The system geometry has been parameterised using axially-symmetric cylindrical coordinates as $(h, r)$. We employ the following hyperbolic geometry to describe the shape of the tip,

$$(h, r) = (h + h_0, h\, tanh(r/2r_0)), \qquad (1)$$

where $h_0$ is the tip height above the surface and $r_0$ is the effective radius of the tip apex. This parameterisation has been adopted because the tip apex is locally spherical, and it has previously demonstrated success in modelling the optical properties of STM tips. We assume that the tip is an ideal metal with $r_0 = 20$ nm and $h_0 = 0.6$ nm. Note that the exact dimensions of the experimental tip are unknown and the values of $r_0$ and $h_0$ are estimated. However, their precise value within expected ranges (10 nm < $r_0$ < 50 nm, 0.5 nm < $h_0$ < 1 nm) have negligible influence on STM modelling. This is because in the expected limit $h_0 \ll r_0$, the tip apex is effectively flat on the scale of a surface point defect positioned directly beneath it (which has a width of approximately 0.3 nm).

The tip is positioned above a flat plane which represents the surface of the boron-doped diamond. The bulk region of the sample has an acceptor concentration of $10^{18}$ cm$^{-3}$ with an appropriate space charge density as discussed below. The posterior surface of the sample represents the metallic backplate and has been separated from the bulk region using an infinite element domain. A graphical representation of the system geometry is presented in Figure SI1.

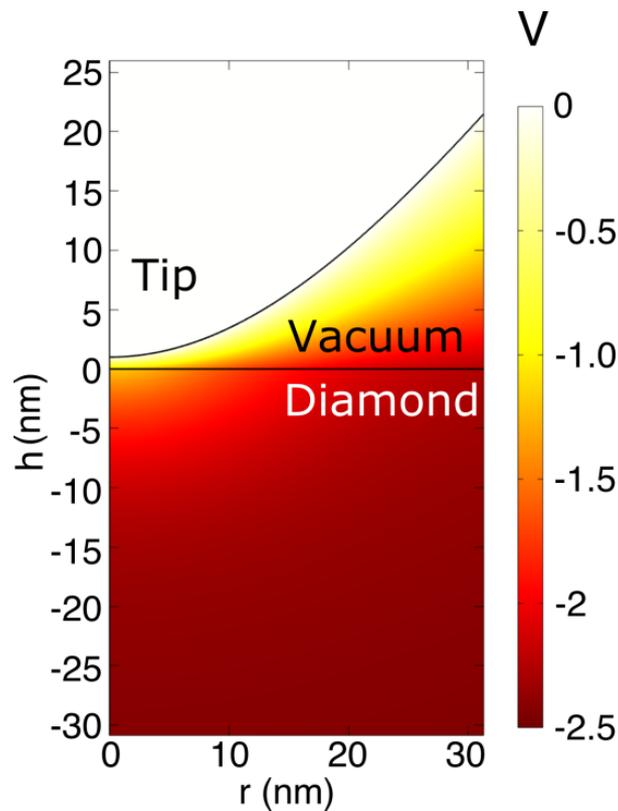

**Figure SI1:** COMSOL simulation demonstrating the electrostatic modelling of an STM tip above boron-doped diamond. The geometry is parameterised using axially-symmetric cylindrical coordinates $(h, r)$. The figure therefore depicts a 2D slice of the 3D geometry taken through the centre of the tip. Note that the bulk diamond region extends down to $h = -100$ nm and across to $r = 100$ nm within the simulation. These full dimensions are not displayed for compactness. In the $-h$ direction, an infinite element then domain separates the bulk region to the sample backplate. Dirichlet boundary conditions on the tip surface and backplate then define the sample-tip bias. In this example, a bias of -2.5 V has been chosen. The heat map displays the electrostatic potential throughout the system as determined by the solution to Poisson's equation. Further details are provided in Chapter 4 of [3].

The space charge density within the sample is described as
$$\rho = \rho_h + \rho_A^-.$$
The density of holes, $\rho_h$, is defined as
$$\rho_h(\varphi) = +\frac{2}{\sqrt{\pi}} q N_V F_{1/2}\left(\frac{(E_{\text{VBM}} - q\varphi) - \mu}{k_B T}\right),$$

where $\varphi$ is the electrostatic potential, $q$ is the electron charge, and $N_V \approx 10^{19}$ cm$^{-3}$ is the effective density of valence states at room temperature ($K_B T \approx 25$ meV). The sample Fermi level, $\mu$, is taken relative to the energy of the valence band maximum ($E_{\text{VBM}}$). Its value of $\mu = 0.194$ eV can be determined in the usual manner through charge conservation. The function $F_{1/2}$ is the complete Fermi-Dirac integral given by,

$$F_{1/2}(x) = \int_0^\infty \frac{\sqrt{y}}{1 + \exp(y - x)} dy.$$

Note that the electron density is negligible and can be neglected. The density of ionised acceptors is given by

$$\rho_A^-(\varphi) = \frac{-qN_A}{1 + 2\exp\left(\frac{(E_A - q\varphi) - \mu}{k_B T}\right)},$$

for $N_A = 10^{18}$ the concentration of substitutional boron defects, and $E_A = 0.37$ eV is the energy of the boron acceptor level relative to $E_{\text{VBM}}$.

Poisson's equation is then solved using finite element analysis. The initial conditions are defined by the applied bias ($V_b$) between the tip surface and the sample backplate (both assumed to be ideal metallic). The solution yields the electrostatic potential throughout the combined tip and sample system. Only the potential directly beneath the tip apex (i.e., $(0, h)$ in the parameterisation of equation (1)) is relevant to the STM simulations. The electrostatic potential along this line is then integrated into the first-principles simulations as discussed below. For example, the voltage drop ($V_d$) is calculated as the difference in potential between the tip apex and the position on the sample surface directly beneath this apex. For a sample bias of $V_b = +2$ V the voltage drop is found to be $V_d = +1.6$V.

### 1.2.2 First-principles simulations

First-principles calculations were performed using the Vienna *ab initio* software package (VASP) [5–7]. The H–C(100): $2 \times 1$ surface was represented using a 420-ion slab with a thickness of 8 carbon layers. The lateral dimensions were 17.7 Å $\times$ 15.2 Å and a graphical representation is provided in Figure SI2. For each defect, two sets of calculations are performed using PBE [8] and HSE06 functionals [9]. The PBE calculations are used to obtain the wavefunctions employed for simulating the STM topographies, whereas the HSE06 calculations are used to accurately determine the one-electron energies of unoccupied states.

First consider the PBE calculations. The slab geometries (including defect) are optimised to a force tolerance of $5 \times 10^{-3}$ eV/Å per ion whereas the total energy is optimized to $10^{-4}$ eV. A planewave cut-off of 600 eV is used and the *k*-point mesh is sampled at the Γ-point. A constant electric field is also included in the simulations self-consistently. The value of the field is determined by the previous electrostatic modelling through the method prescribed in Chapter 4 of [3].

A single-shot calculation is then performed using a $4 \times 4 \times 1$ Monkhorst-Pack *k*-point mesh [10] and the optimised geometry. The resulting wavefunctions must then be modified by the electrostatic potential generated by the tip and sample space-charge density. This is achieved using perturbation theory and the methodology presented in [3]. These perturbed wavefunctions are then subsequently employed for the simulation of STM topographies.

The PBE functional employs the generalised gradient approximation and therefore suffers from the "band gap problem"; a systematic underestimation of the energies associated with unoccupied molecular orbitals. This presents an impediment to STM simulations as electronic energies must be accurately aligned relative to the tip Fermi level. Fortunately, the band gap problem can largely be overcome with the use of hybrid functionals (but at the cost of computational efficiency). For example, the HSE06 functional provides good description of the bulk diamond band gap [3] and diamond surface states [11]. It was therefore employed in this study to determine the energy of unoccupied defect states.

HSE06 calculations were performed for each defect using the PBE-optimised geometries. The computational parameters were identical to the PBE calculations except that the *k*-point mesh was only sampled at the Γ-point. As expected, the energies of the unoccupied levels in the HSE06 calculations were approximately 1–2 eV higher than those of the PBE calculations. However, analysis of their respective charge densities identified no significant differences. The energies of the unoccupied PBE wavefunctions were therefore increased to the HSE06 energies for the simulation of STM topographies. The primary effect was that many unoccupied orbitals were raised in energy above the tip Fermi level, and therefore did not contribute to tunneling at positive sample biases.

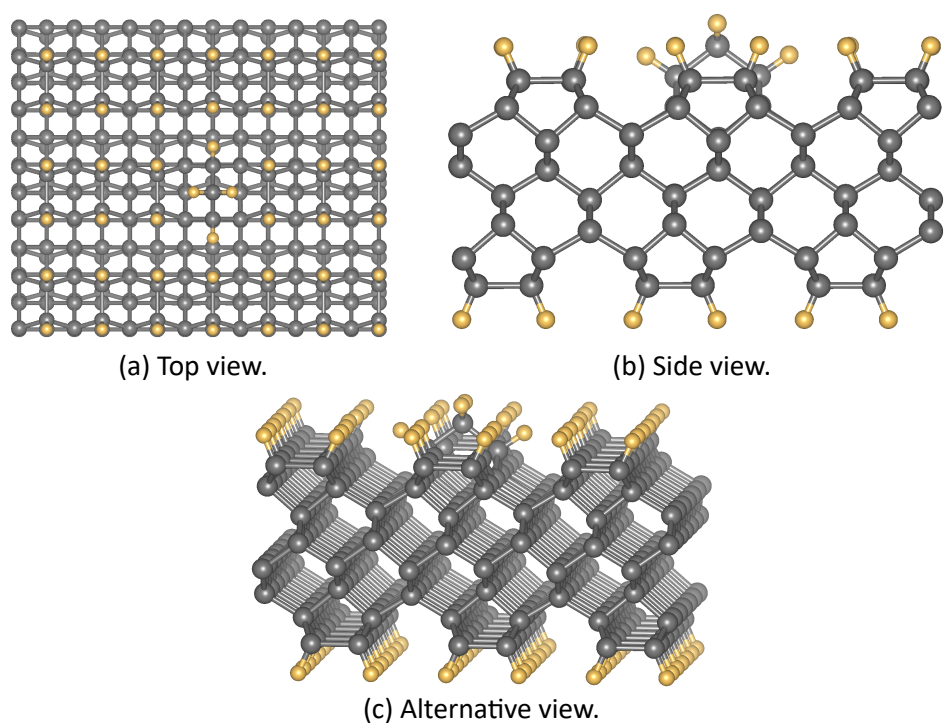

(a) Top view.  (b) Side view.

(c) Alternative view.

**Figure SI2:** Geometry of the 420-ion slab used to represent defects on the H–C(100): $2 \times 1$ surface. It consists of eight carbon layers (grey spheres) with hydrogen termination (gold spheres) on both surfaces. The geometry is periodic in the lateral dimensions. Defects have been positioned in the centre of the slab, and the bridge defect (X of Figure X) has been depicted as a representative example.

### 1.2.3 Simulation of STM topographies

The perturbed PBE wavefunctions were then used to simulate STM topographies. This was performed using conventional means for STM simulation. Specifically, the tunnelling matrix elements were calculated using Chen's approximation [12] to Bardeen's perturbation theory [4]. The tip electronic structure was assumed to be a pure $d_{3z^2-r^2}$ orbital with a density of states of 0.05 eV$^{-1}$. The difference in work function between tip and sample was taken to be 0 V. Simulations were performed using an in-house Fortran code.

Note that these parameters for the tip electronic state are estimates. While they are consistent with expectations for an Ag-coated tip, the true values are specific to a given tip geometry (which can even vary during imaging). They are therefore difficult to determine *a priori*. Nonetheless, these parameters were found to provide an excellent fit to most of the experimental topographies. We specifically note that additional fitting tailored to each defect *was not performed*. That is, the same set of tip parameters were used for all defect simulations.

## 2. STM defect library

### 2.1 Defect classification

STM images of diamond surface defects were simulated using the multi-scale model and classified based on their symmetry. These have been compiled into a 'defect library' which is presented in Figure SI3. Each image has been simulated at a voltage of +2 V and a constant current of 200 pA. A total of 25 defects were considered which comprise a variety of structures, elements, and bonding chemistry expected for as-grown H–C(100): $2 \times 1$ surfaces. These include adsorbates expected to form during CVD [13], common defects found on other semiconducting surfaces [14], and impurities introduced by the STM tip.

The STM topographies were classified in four different ways depending on their symmetry. Moreover, these symmetry classes are directly related to the electronic structure of the defects. The different symmetries are labelled as $1s^+$, $1s^-$, $2s$, and $2p$, and have been assigned visually based on the qualitative features of each image. These qualitative features were discussed in the main text but are briefly re-stated here for convenience. They are: 1s+, a singular bright feature; $1s^-$, a prominent and singular dark feature; 2s, a doughnut shaped/ring feature; and 2p, two bright features separated by a nodal plane. Note that these labels refer explicitly to the symmetry of the STM topography and are not necessarily representative of the defect's orbital structure. However, as discussed in the main text, these symmetries reflect electronic and magnetic properties of the defects.

The four-way classification system encapsulates the properties of all defects considered in Figure SI3. Some defects produce topographies which are best described through a combination of symmetry classes. For example, the C2H bridge (xxiv) hosts a dangling bond with $sp^3$ hybridisation, but due to its geometry produces an STM image with both $1s^+$ and 2p character. A more robust classification system could be obtained through quantitative fitting of the topographies to cross sections of spherical harmonics. However, this was considered unnecessary for the purposes of this work. While the 25 defects considered here are not exhaustive, they are representative of most electronic structures and bonding geometries expected on as-grown H–C(100):2 × 1 surfaces. Hence, the symmetry classification should offer decent predictive power for the electronic properties of other defects.

### 2.2 Systematic overestimation of dark features

The overestimation of dark features occurs consistently throughout the simulations. For example, consider the simulated and experimental topographies for the 2s defects in Figure 3 (c) of the main text. The line profiles reveal that the multi-scale model correctly reproduces the lateral width of the ring features. However, their depth is overestimated by up to 1 Å.

This discrepancy is likely attributable to the bluntness of experimental tips. The multi-scale model employs Chen's approximation to calculate the tunnelling matrix elements [12]. Chen's approximation assumes that all tunnelling current flows via a single atom at the tip apex. This assumption is somewhat unrealistic, and experimental tips likely contain a blunt cluster of atoms which form their apex. The total current can therefore be approximated as the sum of the tunnelling currents via each atom in the cluster. However, the tunnelling amplitude scales exponentially with distance. Hence, the current measured above dark features will be overestimated due to erroneous tunnelling via other surface states proximate to the cluster. Following the same reasoning, bright features are much better represented using Chen's approximation. This is because the tunnelling current is dominated by the tip atom closest to the defect.

Ultimately, a better fit to the experimental line profiles can be obtained by averaging the simulated tunnelling current over an effective radius which represents the bluntness of the tip. This is demonstrated explicitly for the $CH_2$ bridge defect ((xv) of Figure SI3) which was identified in Figure 3 (b) of the main text. The overestimation of dark features was remedied through a spatial convolution of the tunnelling current with a 3D Gaussian. A variance of 1 Å in each dimension is chosen to reproduce the effective bluntness of the tip. This value is physically reasonable and is approximately the Wigner-Seitz radius of an Ag atom. The resulting line profile (in the direction parallel to the dimer rows) is indicated in Figure 3 (b) of the main text through a dashed blue line. It is also reproduced and enlarged below in Figure SI4 for convenience. Convolution successfully reduced the discrepancy in corrugation amplitude between simulation and experiment from 0.5 Å to 0.15 Å. While some disagreement remains, this effectively demonstrates how bluntness of the tip apex can reduce the apparent depth of dark features.

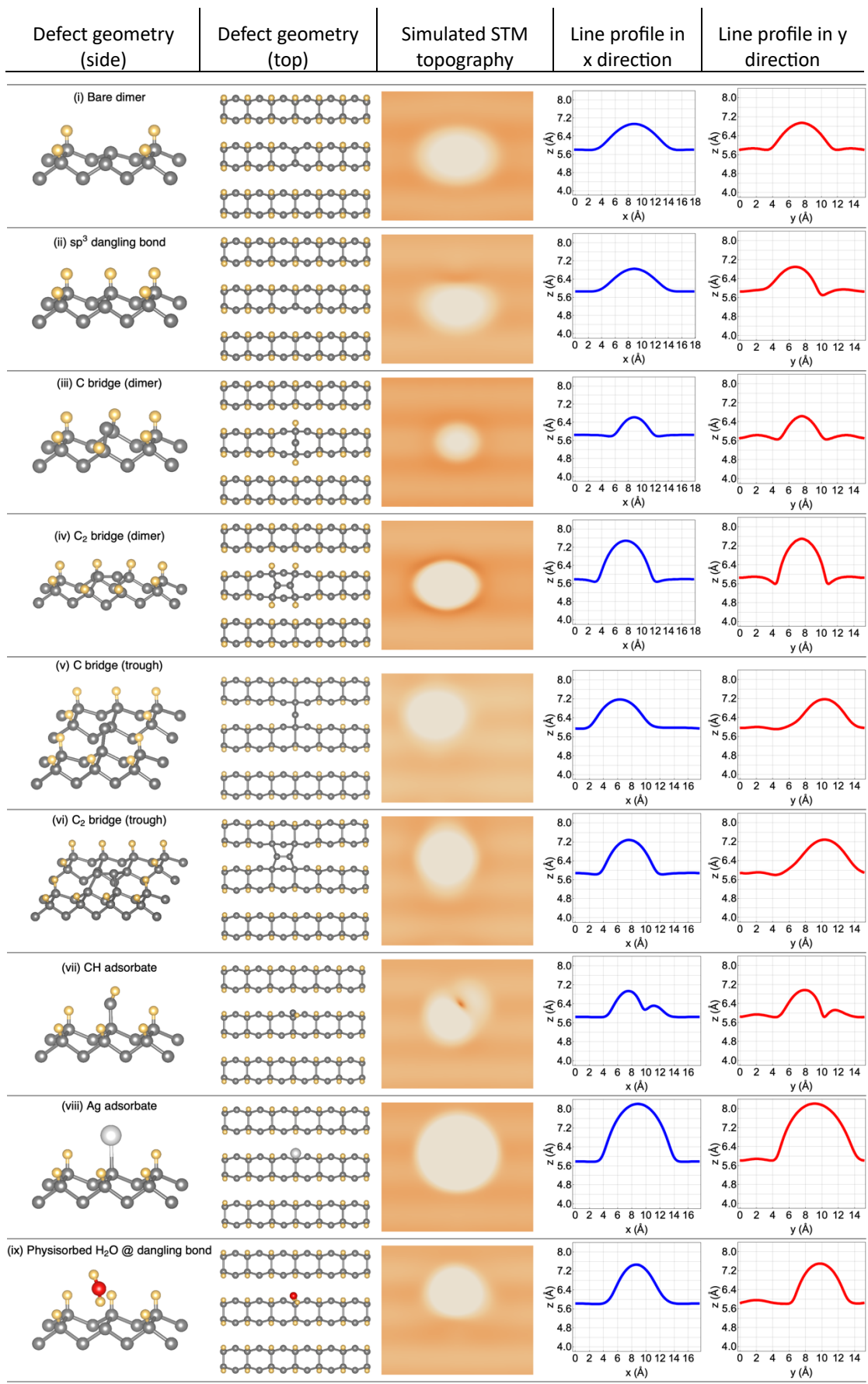

(a) 1s$^+$ defects (Figure SI3 continues overpage).

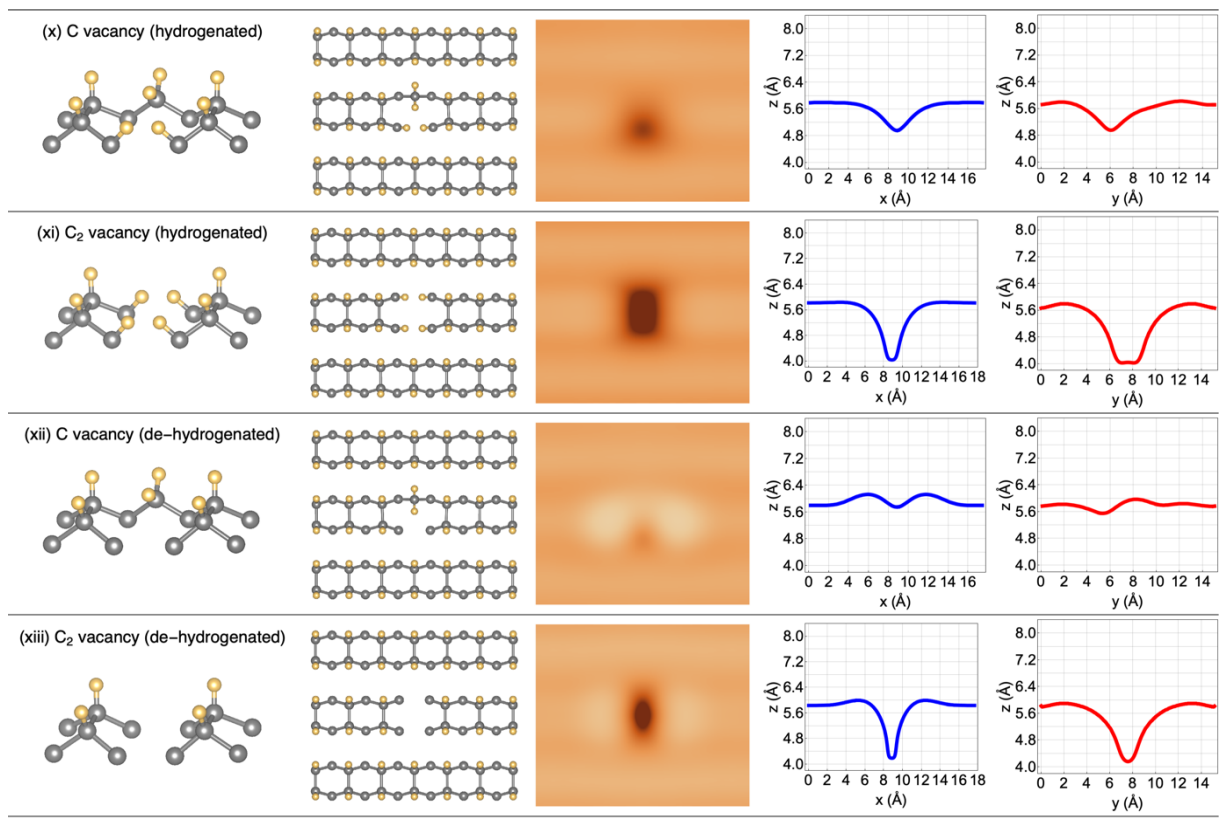

(b) 1s- defects (Figure SI3 continues overpage).

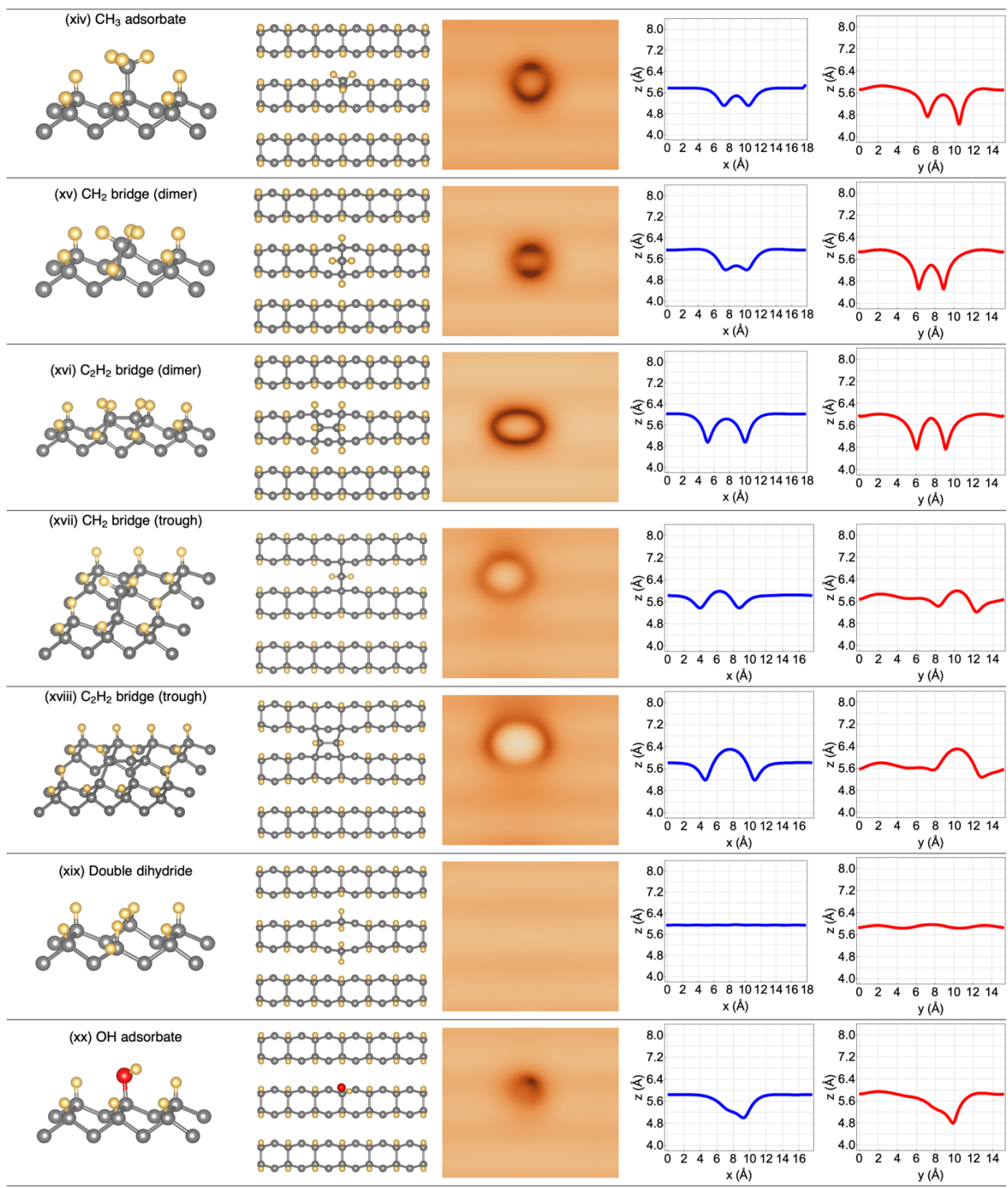

(c) 2s defects (Figure SI3 continues overpage).

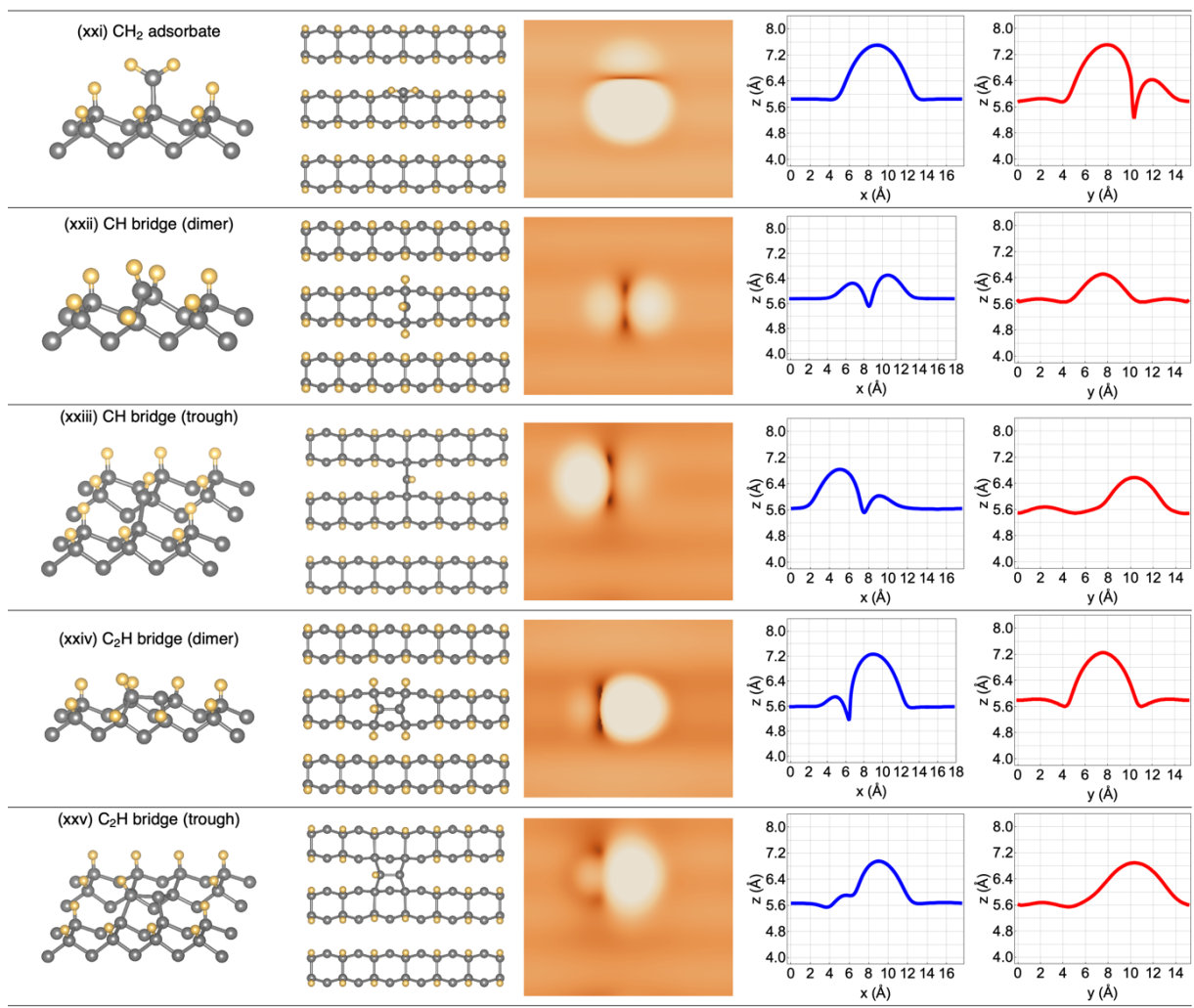

(d) 2p defects.

**Figure SI3:** Defect library for the H–C(100):2 × 1 surface. Geometries and constant current STM images are simulated for 25 different defects. These assume I = 0.2 nA, $V_{bias}$ = 2 V, and an Ag tip for consistency with the experimental results. One-dimensional line profiles have been sampled through the centre of each defect and run either parallel or perpendicular to the dimer rows. Defects have been classified into four symmetry groups based on the qualitative features of their STM topographies. These are $1s^+$ (a single bright feature), $1s^-$ (a prominent dark feature), 2s (a 'doughnut' feature), and 2p (two bright features separated by a nodal plane). As discussed in the main text, this symmetry is associated with the defect's electronic and magnetic properties.

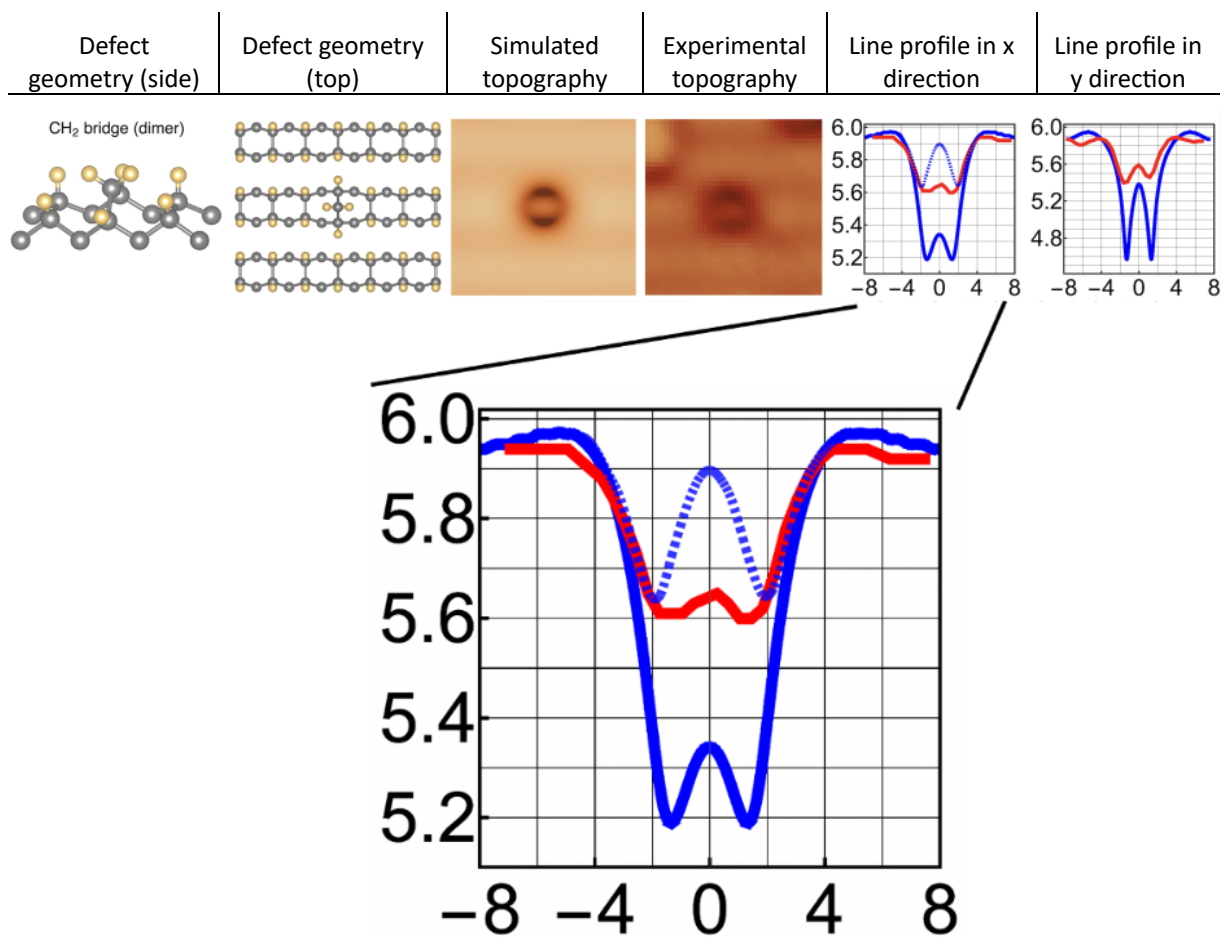

**Figure SI4**: Reproduction of Figure 3 (b) from the main text which demonstrates identification of the CH$_2$ bridge defect. The line profiles in the x direction have been magnified to highlight the overestimation of dark feature depth. Note that all units are in angstroms. The simulated line profile (blue solid line) reaches a depth of approximately 5.2 Å, whereas the experimental line profile only reaches 5.6 Å. We suspect that this discrepancy is due to the bluntness of the tip apex. To validate this hypothesis, convolution with a Gaussian was used to 'average' the tunnelling current within a sphere of 1 Å about the tip apex. The resulting line profile is depicted using a dashed blue line, which reaches a depth of approximately 5.6 Å. While this is consistent with the depth of the experimental line profile, the height of the 'hole' in the donut feature remains inadequately represented. Nonetheless, this demonstrates that the overestimation of dark features can likely be attributed to the bluntness of the experimental tip apex.

### 2.3 Electronic and magnetic properties of observed defects

The electronic band structure for the defects observed on the diamond surface (i.e., those presented in Figure 3 of the main text) are presented in Figure SI5. While the band structures have been calculated using PBE functionals, the energy of unoccupied states have been raised by 1.54 eV to provide a better representation of the true surface band gap and defect energies[151]. Ideally, hybrid functionals would be used to produce the band structure, but this is not possible due to computational constraints associated with the large size of the supercells (~420 ions).

Note that the *k*-space samples the reciprocal space of the supercell, not the primitive 2 × 1 surface unit cell. Red dots represent one-electron states which are associated with the defect structure. The results have been separated by symmetry class, and the examples shown can be considered representative of most other defects within the same symmetry class.

The defect free surface is presented in Figure SI5 (a). It contains all features expected of the H–C(100):2 × 1 surface, notably the presence of parabolic bands centred at the Γ point with an energy of approximately 5 eV above the VBM. These are the image states associated with the surface's negative electron affinity and possess wavefunctions which extend beyond the diamond surface[16,17].

The band structures for the $1s^+$ defects are presented in Figure SI5 (b). The most prominent features are flat bands positioned several hundred meV above the surface VBM and between 3 and 4 eV above the surface VBM. These are the occupied bonding orbital and unoccupied anti-bonding orbital associated with the defect. Analysis of their respective charge densities (not presented) reveal that these states are highly localised to the defect. At the sample biases of 2–2.5 V used in this study, the bonding orbital contributes strongly to the STM current, producing the prominent bright feature observed in the STM topographies all $1s^+$ defects. The anti-bonding orbital does not contribute to the STM current at this voltage because it is too high in energy.

For *n*-type samples (which are typically doped using N or P donors), the sample fermi level resides above the anti-bonding orbital. These states therefore act as charge traps which are occupied in the absence of an applied bias. However, in the presence of an electric field, charge may rearrange amongst these surface defects thereby producing electrical noise or surface screening. This screening is generally undesirable because it prevents external electric fields from penetrating into the diamond substrate. Consequently, near-surface NV centres are unable to detect these external fields.

In contrast, the $1s^-$ and 2s defects do not possess any isolated mid-gap states. This can be viewed in Figure SI5 (c) and (d), which reveals that defect-related states are dispersed throughout the valence band. No significant defect-related states exist near the VBM, and therefore these defects do not produce a prominent bright feature in the STM topography. Instead, the depressive features result from an absence of any defect-related states amongst the other states of the H–C(100):2 × 1 surface.

The band structures for the 2p defects are presented in Figure SI5 (e). Like the $1s^+$ defects, these band structures possess defect-related bonding and anti-bonding orbitals within the band gap. The bonding states may be conceptualised as unhybridised 2p orbitals. These produce the bright lobes which are prominent in the STM topographies.

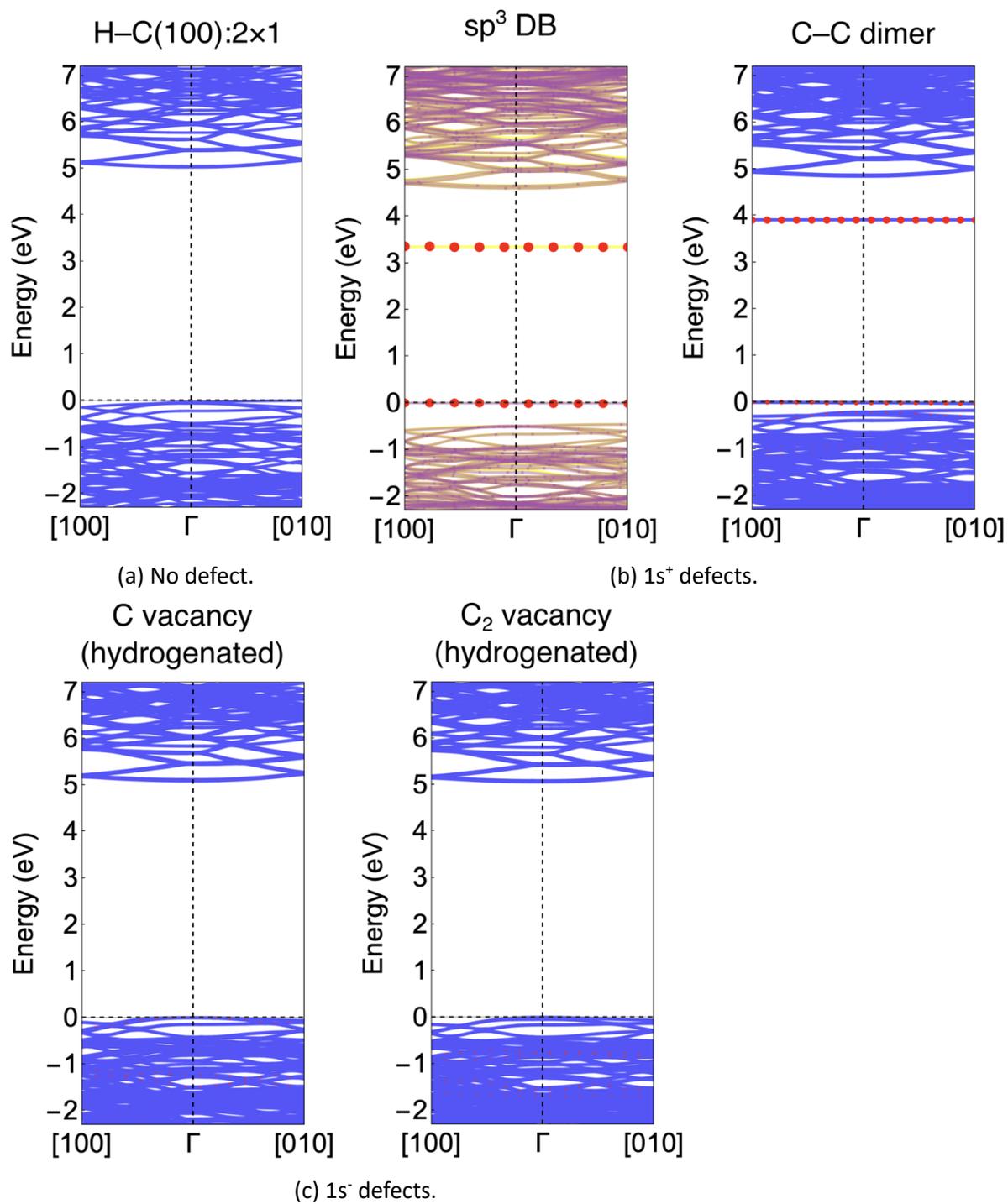

**Figure SI5**: Continued overpage.

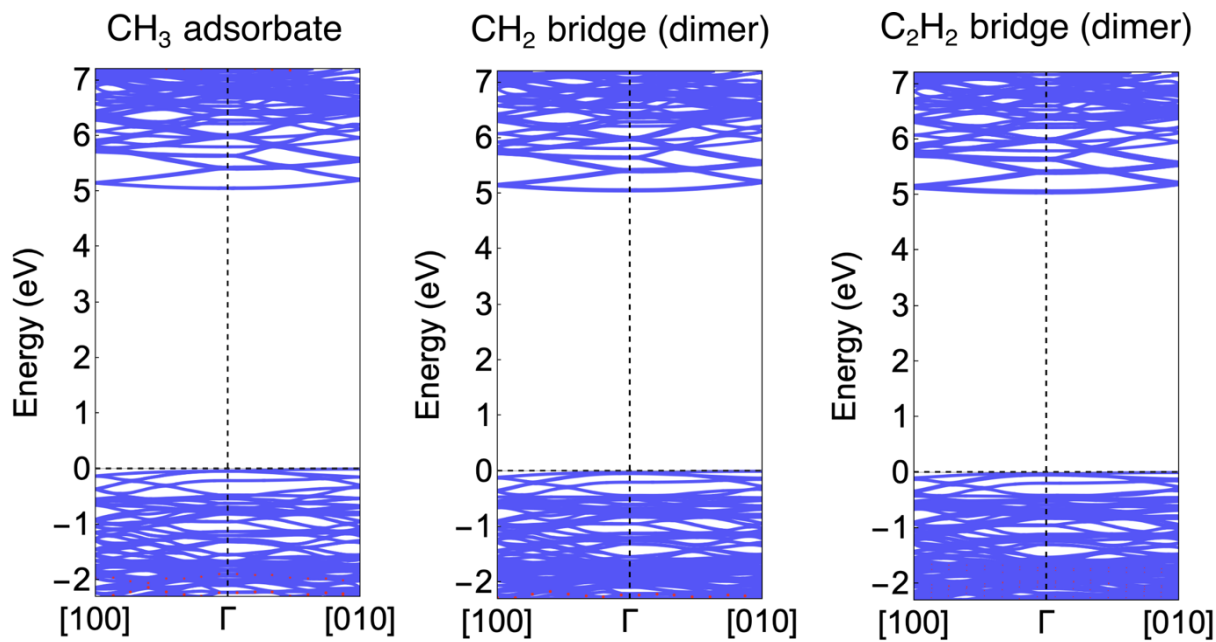

(d) 2s defects.

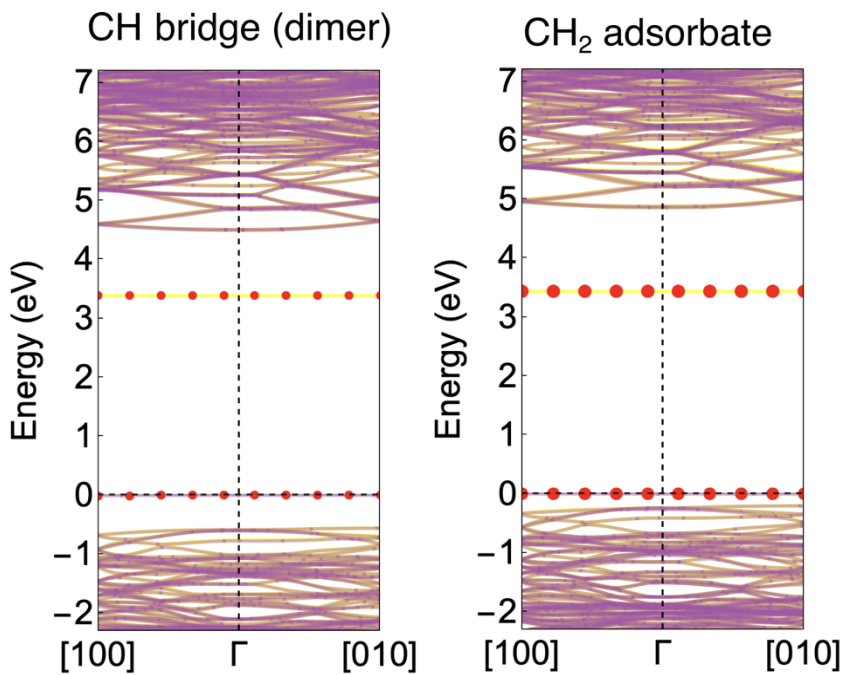

(e) 2p defects.

**Figure SI5**: Electronic band structures for defects observed on the diamond surface. These correspond to the defects presented in Figure 3 of the main text. The energy scale is relative to the energy of the highest occupied molecular orbital. The *k*-space of the supercell is sampled from the [100] high-symmetry point (in the direction parallel to the dimer rows) to Γ, and then from Γ to the [010] high-symmetry point (in the direction perpendicular to the dimer rows). Band structures in blue are calculated without spin polarisation, while those with purple bands (spin down channel) and yellow bands (spin up channel) are calculated with spin polarisation. Note that defect related states have a characteristically flat band structure. Furthermore, the red dots represent one-electron states which are associated with the given defect. This has been calculated through projection of the charge density about Wigner-Seitz cells that are centred on each ion which comprises the defect. The size of the red dot provides a rough representation for the magnitude of this projection.

## 3. Additional examples of experimental defects

Figures SI6, SI7, and SI8 demonstrate further experimental topographies of the H–C(100):2 × 1 surface. Figure SI6 presents a large scale STM image demonstrating a variety of surface defects. The majority of these defects are the $CH_2$ bridge, indicated by the presence of a "doughnut" feature positioned in the centre of the dimer row. These can be differentiated from other 2s defects which possess a ring feature that is "off-centre" from the dimer row. Two examples, likely $CH_3$ adsorbates, are highlighted by white squares.

Figure SI7 presents an STM image with both C and $C_2$ vacancies. These can be easily differentiated by the width of the dark feature as indicated.

Figure SI8 similarly presents an STM image with both an $sp^3$ DB and a C–C dimer. These can be differentiated through two means. Firstly, the bright feature of the C–C dimer is notably larger than that of the $sp^3$ DB. Secondly, the $sp^3$ DB is not symmetric with respect to the underlying dimer rows.

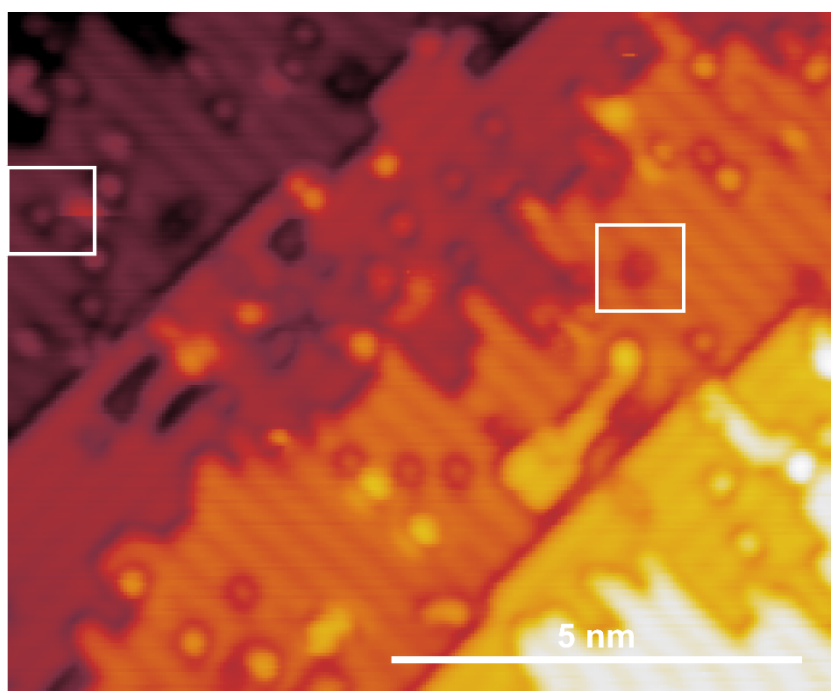

**Figure SI6:** A large scale STM image demonstrating a variety of surface defects. The majority of these are 2s defects, indicated by the presence of a ring or "doughnut" feature. Most of these are $CH_2$ bridge defects, which are symmetric about the centre of the dimer row. These can be differentiated from $CH_3$ adsorbates which are positioned on either side of the dimer row. Two such examples have been highlighted using white squares. Tunnelling parameters: V = 1.5 V; I = 200 pA.

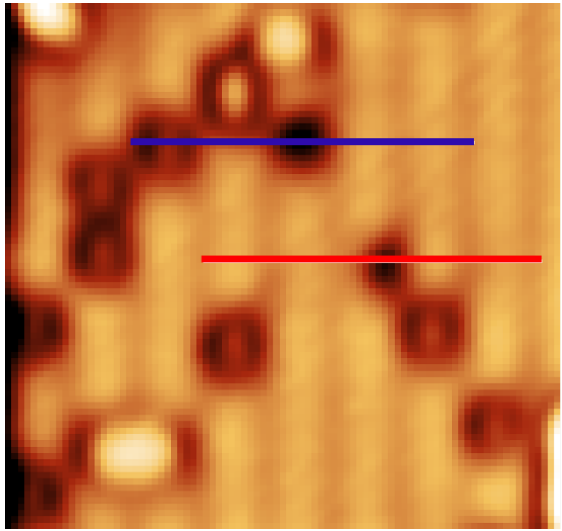 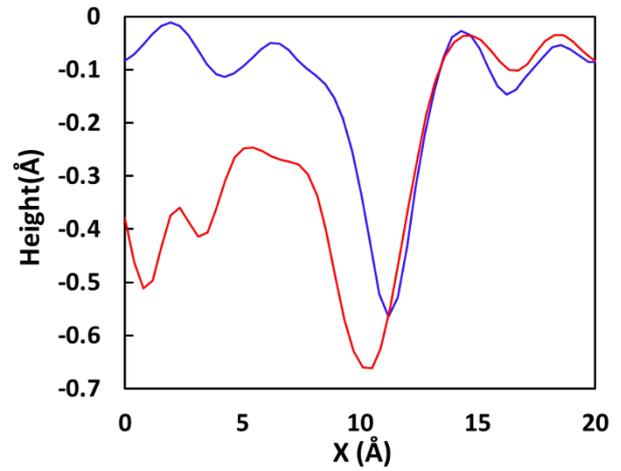

**Figure SI7:** Line profiles show the width of the C and $C_2$ vacancies. The tip trajectory for line profiles across the C and $C_2$ vacancies are illustrated with red and blue lines, respectively. Tunnelling parameters: V = 2.5 V; I = 200 pA.

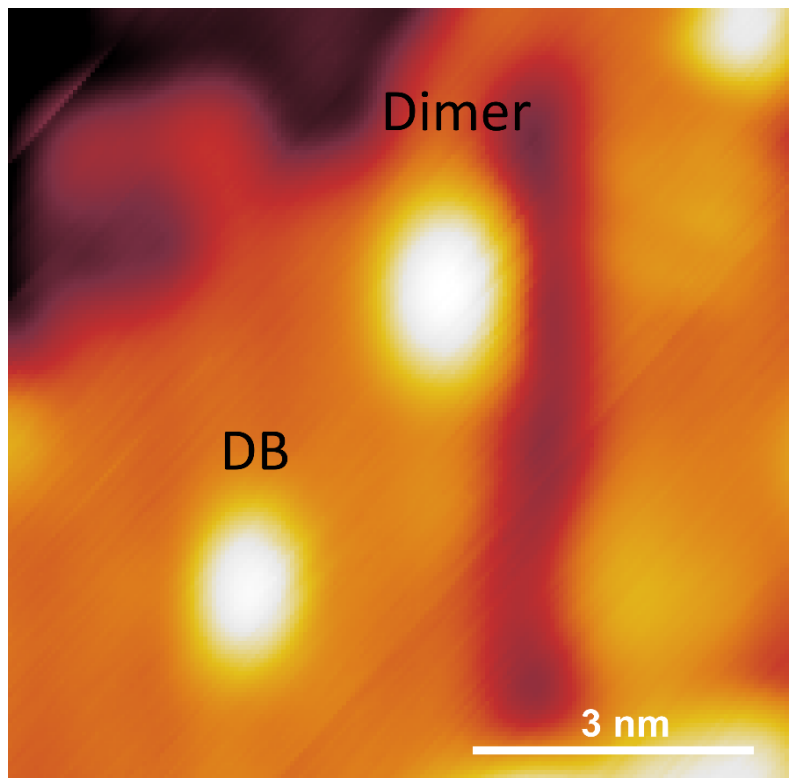

**Figure SI8:** The STM image shows an $sp^3$ DB, and a bare C–C dimer. Tunnelling parameters: V = 2.0 V; I = 200 pA.

# 4. Diamond overgrowth mechanism during CVD

This section aims to give further context regarding accepted diamond CVD growth processes. This model has been established through decades of first-principles simulations [13], mesoscale modelling of growth rates and surface morphology [18,19], and plasma experiments and simulation [20]. In the following, we describe this accepted model through DFT simulations of the overgrowth reaction. We find that the stable intermediaries are the same surface defects identified on the as-grown surface using STM.

## 4.1 Computational details

The geometries of all intermediates and products were optimized using VASP at GGA-PBE level of theory. The slab size is smaller than that used for the STM simulations, with dimensions of $3 \times 2 \times 4$ in terms of primitive diamond unit cells. The energetic effect of all structures entering ($CH_3$, H) or leaving ($H_2$) during the multi-cascade chemical transformation was taking into account by separate optimization of their geometries and subtracting obtained energies from the energy of diamond slab. Transition states were identified using Nudged Elastic Band (NEB) calculations.

## 4.2 Overgrowth mechanism

Figure SI9 depicts elementary steps of the overgrowth mechanism occurring on the H–C(100):$2 \times 1$ surface. It consists of 5 successive reactions. First, a H radical from the plasma abstracts a H atom from the diamond surface. This produces a reactive DB. Next, an incoming $CH_3$ radical binds to this DB and forms the stable intermediate **3** allocated at -3.71 eV on the energy profile. Note that this is the $CH_3$ adsorbate identified in (xiv) of Figure SI3 (c). The adsorbed $CH_3$ then converts to $CH_2$ via a second abstraction of a H radical. The mechanism continues by expanding the 5-membered ring of the underlying dimer into 6-membered via incorporation of $CH_2$ group into the diamond lattice. The cascade finishes with adsorption of a H radical and the formation of energetically stable $CH_2$ mono-bridge **6**. Note that this is the same defect identified in (xv) of Figure SI3 (c).

Site **6** then acts as a nucleation point for the next layer of diamond growth. A reaction cascade similar to that presented in Figure SI9 results in another $CH_2$ bridge defect positioned on an adjacent dimer. As shown in Figure SI10, this pair of $CH_2$ bridges may merge to produce the $C_2H_2$ bridge absorbate **8**. This is the same defect identified in the STM topographies of the main text and corresponds to (xvi) in Figure SI3 (c).

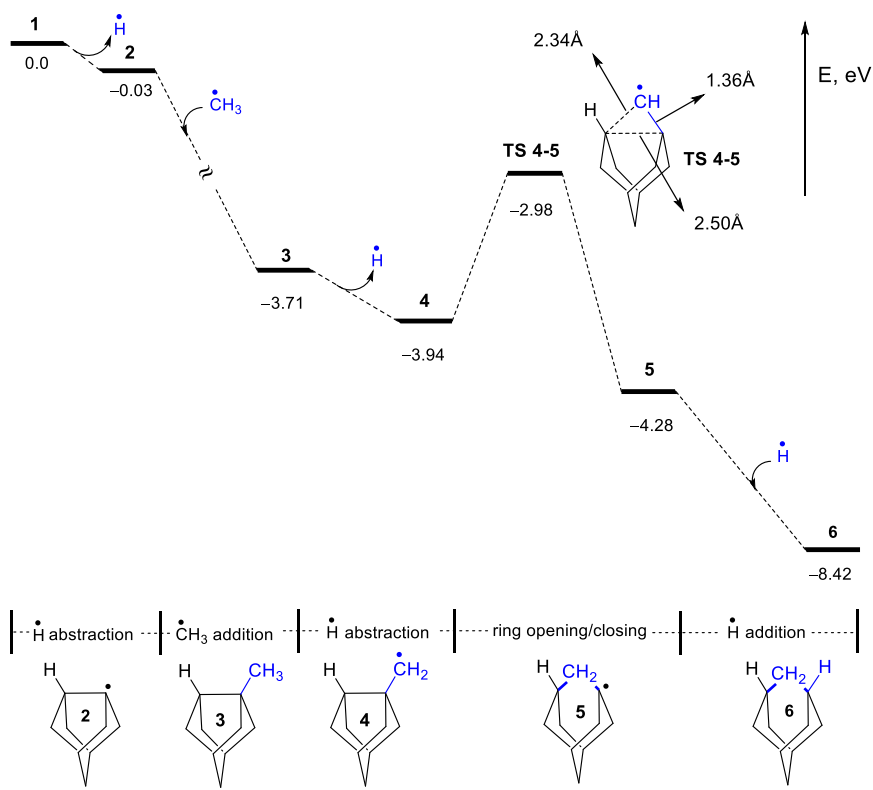

**Figure SI9:** Reaction path describing the formation of a CH₂ bridge defect during CVD overgrowth. The lower diagrams depict the structure of a single dimer of the H–C(100):2 × 1 surface throughout the reaction cascade. The upper diagram displays the relative energy of each structure (in eV). Structures and their energies have been calculated using VASP with PBE functionals.

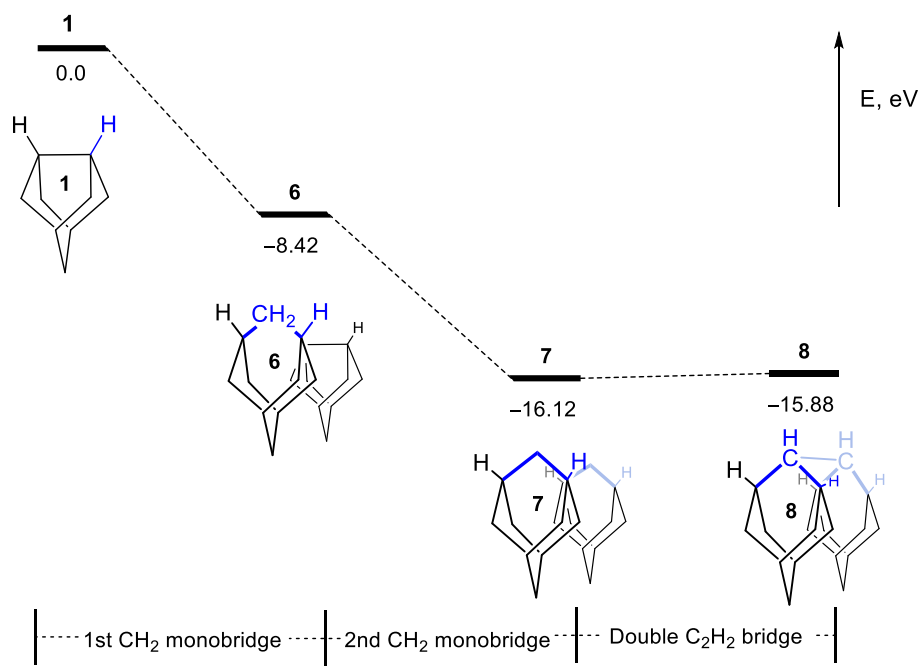

**Figure SI10:** Reaction path describing the formation of a C₂H₂ bridge defect during CVD overgrowth. Process 1–6 describes a similar reaction to that describe in Figure SI8. This produces a second CH₂ bridge defect along a dimer adjacent to a first CH₂ defect. Two successive H abstractions create DBs on each of the CH₂ bridge defects. These defects then merge through a covalent bond to form a C–C dimer. This is precisely the C₂H₂ bridge defect, which acts as a stable site for nucleation of the next diamond layer. Structures and their energies have been calculated using VASP with PBE functionals.